\documentclass[useAMS,usenatbib]{mn2e}
\usepackage[dvips]{graphicx}
\usepackage{amsmath}
\usepackage{amssymb}
\usepackage{multirow}
\usepackage{mathrsfs}
\usepackage[active]{srcltx}   

\usepackage[active]{srcltx}   

\title[A new model for the full shape of $P(k)$]{A new model for the full shape of the large-scale power spectrum}
\author[F. Montesano et al.]
{Francesco Montesano\thanks{E-mail: montefra@mpe.mpg.de}, 
Ariel G. S\'anchez and Stefanie Phleps\\ 
Max-Planck-Institut f\"ur Extraterrestrische Physik, Giessenbachstra{\ss}e, 85748 Garching, Germany.\\
}

\begin{document}

\date{Accepted xxx. Received xxx; in original form xxx}

\pagerange{\pageref{firstpage}--\pageref{lastpage}} \pubyear{0000}

\maketitle

\label{firstpage}

\begin{abstract}

We present a new model for the full shape of large-scale the power spectrum based on renormalized perturbation theory.
To test the validity of this prescription, we compare this model against power spectra measured in a suite of 50 large volume, moderate
resolution N-body simulations. Our results indicate that this simple model provides an accurate description of the full shape of
the power spectrum taking into account the effects of non-linear evolution, redshift-space distortions and halo bias for scales $k\lesssim0.15 \,h\,\rm{Mpc^{-1}}$, making
it a valuable tool for the analysis of forthcoming galaxy surveys. 
Even though its application is restricted to large scales, this prescription can provide tighter constraints on the dark energy equation of state parameter $w_{\rm DE}$
than those obtained by modelling the baryonic acoustic oscillations signal only, where the information of the broad-band shape of the power spectrum  is discarded. 
Our model is able to provide constraints comparable to those obtained by applying a similar model to the full shape of the correlation function, which is affected by
different systematics. Hence, with accurate modelling of the power spectrum, the same cosmological information can be extracted from both statistics.

\end{abstract}

\begin{keywords}
large-scale structure of Universe -- cosmology: theory.
\end{keywords}

\section{Introduction}

Over the past decade, new cosmological observations have revolutionised our view of the Universe by showing increasing evidence that it is currently undergoing a phase of accelerated expansion. The Hubble diagram of Type 1a supernovae (SN1a) \citep{riess_98, perlmutter_99, kowalski_unionSN1a}, together with other independent data sets like observations of the temperature fluctuations in the cosmic microwave background (CMB)  \citep{hinshaw_wmpa1_aps,Spergel_03,Spergel_07,komatsu_wmap09} combined with measurements of the large scale structure (LSS) of the galaxy distribution \citep{efstathiou02,percival02,tegmark04,Sanchez_06, Sanchez_09,percival_09_SDSS,reid_10_SDSS}, have  established the picture of a geometrically flat Universe whose energy content is composed of about $4\%$ baryonic matter, $21\%$ of cold dark matter and $75\%$ of a mysterious component called Dark Energy (DE), responsible for the current acceleration of the expansion rate. The nature of dark energy is one of the most important open questions in modern cosmology as the comprehension of its nature has deep implications for our understanding of fundamental physics. In the past few years a substantial effort has been devoted to constrain the properties of dark energy as accurately as possible in order to distinguish between possible alternative descriptions of this phenomenon \citep[see][for a review]{copeland_DErev}. This effort will continue in the coming years with the construction of the next generation of large galaxy surveys, like the Panoramic Survey Telescope \& Rapid Response System \citep[Pan-STARRS,][]{panstarrs}, the Dark Energy Survey \citep[DES,][]{des}, the Baryonic Oscillation Spectroscopic Survey \citep[BOSS,][]{schlegel_BOSS}, the Hobby Eberly Telescope Dark Energy Experiment \citep[HETDEX,][]{hetdex} and, on a longer time scale, the space based Euclid mission \citep{cimatti2008}. These surveys will cover volumes much larger than current datasets, providing a dramatic improvement in the accuracy of the constraints on the values of cosmological parameters, in particular the dark energy equation of state parameter defined as $w_{\rm DE} = p_{\rm DE}/\rho_{\rm DE}$, where $p_{\rm DE}$ and $\rho_{\rm DE}$ denote the pressure and the energy density of this component.

These new surveys will provide a clear picture of the Baryonic Acoustic Oscillations (BAO), a signature imprinted in the large scale galaxy distribution by the acoustic fluctuations in the baryon-photon fluid prior to recombination. These can be seen as a quasi-harmonic series of oscillations of decreasing amplitude in the power spectrum at wave numbers $0.01 \,h\,\rm{Mpc^{-1}} \lesssim k \lesssim 0.4 \,h\,\rm{Mpc^{-1}}$ \citep{Sugiyama_95, Eisenstein_98, EH99}. In the two-point correlation function, the Fourier transform of the power spectrum, the BAO are visible as a unique broad and quasi-gaussian peak \citep{Matsubara_04}. The BAO where first detected in the correlation function of the Luminous Red Galaxies (LRG) sample drawn from the Sloan Digital Sky Survey (SDSS) by \citet{Eisenstein_05} and the power spectrum of the two-degree Field Galaxy Redshift Survey (2dFGRS) by \citet{cole_05_2dF} \citep[see][for more recent analyses]{percival07,cabre_09,gaztanaga08,Sanchez_09,percival_09_SDSS,reid_10_SDSS,kazin_09}. The acoustic scale inferred from the galaxy clustering is related to the sound horizon scale at the drag epoch, i.e. when the baryons where released from the photons. Because of the very high photon to baryon ratio, this epoch happened slightly later that the decoupling epoch \citep{komatsu_wmap09}. As this scale depends solely on the plasma physics after the big bang and can be calibrated using CMB data, it is possible to use the BAO scale as a standard ruler. Measuring the apparent size of the BAO in the directions parallel and perpendicular to the line of sight, it is possible to measure the redshift dependence of the Hubble parameter $H$ and the angular diameter distance $D_{A}$ and thus constrain cosmological parameters like $w_{\rm DE}$ \citep{blake03, Hu_Haiman03, Linder03, Seo_03, Wang06, Guzik_07,Seo_07, Seo_08, shoji_09, seo_09}.

Robust percent level constraints on $w_{\rm DE}$ require measurement of the BAO scale at sub-percent level \citep{nishimichi_scaleBAO, Angulo_08, shoji_09}. In order to realize the full potential of future BAO measurements as a cosmological probe, a detailed analysis of the possible systematic effects introduced by the non-linear evolution of density fluctuations, redshift space distortions and bias is required. Recent studies based on the analysis of large volume N-Body simulations and on novel approaches to perturbation theory \citep[e.g. renormalized perturbation theory, ][]{crocce_RPT1, crocce_RPT2}, have shown that non-linearities introduce shifts of up to few percent in the position of the BAO peaks with respect to the prediction of linear theory, both in the correlation function and in the power spectrum \citep{Smith_07,crocce_nonlinBAO,Sanchez_08}. Redshift-space distortions and bias might introduce additional shifts. When not properly taken into account, these distortions introduce systematic errors in the obtained constraints that can be larger than the expected measurement errors. This highlights the importance of using a model able to take in account these distortions with the accuracy demanded by the future surveys.

The modelling of non-linear distortions in the BAO signal in the power spectrum has been the focus of much theoretical work \citep{angulo_05, Huff_07, Smith_07, Angulo_08, Desjacques_08, Smith_08, crocce_nonlinBAO, Seo_08, seo_09}. Most of these analyses have focused on modelling the BAO signal filtered out from the broad band shape of the power spectrum. This approach attempts to produce a purely geometrical test based on the BAO signal, but it has the disadvantage of discarding useful information contained in the full shape of the power spectrum. \citet{Sanchez_08} showed that the correlation function is much less affected by scale dependent effects than the power spectrum and that its full shape can be described by a simple model, originally proposed by \citet{crocce_nonlinBAO}, with the accuracy required for surveys of up to two orders of magnitude larger than present day samples. \citet{Sanchez_09} applied this model to the LRG correlation function measured by \citet{cabre_09} and performed a detailed analysis of the constraints on cosmological parameters, obtaining an improvement of roughly a factor two in the accuracy of the constraints on the dark energy equation of state with respect to the ones recovered by using only the BAO signal in the power spectrum.

In this work we present a new model for the full shape of the power spectrum which is based on the model for the correlation function of \citet{crocce_nonlinBAO} and \citet{Sanchez_08,Sanchez_09}. We compare this model against the real and redshift space power spectra of the dark matter and halo distributions obtained from an ensemble of large volume, moderate resolution N-body simulations \citep[L-BASICC II][]{Angulo_08, Sanchez_08}. The outline of the paper is as follows: in section \ref{L-PS} we describe the set of simulations used in our work and provide a few technical details about the computation of the power spectrum. In section \ref{model} we introduce our theoretical model of the full shape of the power spectrum. In Section \ref{p_k_fit} we determine the range of scales in which this model is able to accurately describe the results drawn from the numerical simulations and recover unbiased constraints on the dark energy equation of state. Finally section \ref{conclusion} presents our main conclusions.

\section[]{N-body simulations and the \\* computation of the power spectrum}\label{L-PS}

In this section we briefly describe the L-BASICC II ensemble of N-Body simulations used in our analysis \citep[see][for more details]{Angulo_08} and give some technical details about the methodology implemented to compute the power spectra from these simulations. A more detailed description can be found in Appendix \ref{ap:power_spectrum}.

\subsection{The L-BASICC II N-Body simulations} \label{L-Basicc}

We use an ensemble of 50 moderate resolution, very large volume dark matter N-Body simulations called L-BASICC II \citep{Angulo_08,Sanchez_08}. These are analogues of the L-BASICC simulations used in \citet{Angulo_08} and represent the evolution of the dark matter density field in a universe characterised with a flat $\rm\Lambda$CDM cosmology, consistent with the constraints on cosmological parameters obtained from the combination of CMB and LSS information of \citet{Sanchez_06} and \citet{Spergel_07}.  The values of the cosmological parameters and other specifications of the simulations are listed in Table \ref{tab1:sim}.

\begin{table}
 \centering
 \begin{minipage}{80mm}
  \caption{Cosmological parameters and specifications of the L-BASICC II simulations.} \label{tab1:sim}
  \begin{tabular}{llr}
  \hline
  matter density & $\rm{\Omega_{m}}$ & 0.237\\
  baryonic density & $\rm{\Omega_{b}}$ & 0.041\\
  scalar spectral index  & $n_{\rm s}$ & 0.954\\
  Amplitude of density & \multirow{2}{*}{$\rm{\sigma_{8}}$} & \multirow{2}{*}{0.77}\\
   fluctuations &\\
  Hubble constant & $H_{0}$ & $73.5 \,\rm km \,s^{-1}\, Mpc^{-1}$\\
  \hline
   Number of particles & $N_{\rm p}$ & $448^{3}$ \\
   Particle mass & $M_{\rm p}$ & $1.75 \times 10^{12}$ $h^{-1} \rm M_{\odot}$ \\
  Softening length & $\epsilon$ & 200 $h^{-1}$  kpc\\
  Comoving box side &  & 1340 $h^{-1}$ Mpc\\
  Comoving volume & & 2.41 $h^{-3} \rm Gpc^{3}$\\
  Total volume of the ensemble&  & 120 $h^{-3} \rm Gpc^{3}$\\
\hline
\end{tabular}
\end{minipage}
\end{table}

The position and the velocity of all the particles in the simulations were stored in three snapshots at redshifts $z = 0$, 0.5 and 1. Halo catalogues were constructed from the dark matter distributions at each redshift using a Friend-of-Friends (FoF) algorithm \citep{davis_85}, with a linking length parameter $b=0.2$ and selecting all the haloes with more than 10 particles, which corresponds to a minimum halo mass of $1.75 \times
10^{13}$ $h^{-1} \rm M_{\odot}$.

\begin{table}
 \centering
 \begin{minipage}{80mm}
  \caption{Number of haloes ($N_{\rm h}$) and shot noise term ($1/\bar n$) for the total halo sample and the three mass bins defined in section~\ref{L-Basicc} for redshift 0, 0.5 and 1.} \label{tab:bins}
  \begin{tabular}{ c c c c c c c }
   \hline 
    & \multicolumn{2}{c}{$z=0$} & \multicolumn{2}{c}{$z=0.5$} & \multicolumn{2}{c}{$z=1$} \\
    \hline
    bin & $N_{\rm h}$ & $1/\bar{n}$ \footnote{units of $h^{-3}\,\rm{Mpc^{3}}$} & $N_{\rm h}$ & $1/\bar{n}$ & $N_{\rm h}$ & $1/\bar{n}$\\ \hline
    tot & 465903 & 5164 & 294204 & 8178 & 143531 & 16764 \\
    1 & 262232 & 9175 & 151976 & 15832 & 69457 & 34642 \\
    2 & 101825 & 23630 & 71551 & 33628 & 36852 & 65291 \\
    3 & 101846 & 23625 & 70678 & 34043 & 37221 & 64644 \\
    \hline 
  \end{tabular}
 \end{minipage}
\end{table}

From each halo catalogue we extract three sub samples selected according to the following mass limits (in units of $10^{13} h^{-1}M_{\odot}$):
\begin{itemize}
\item[1)]  $M < 3.5$, 
\item[2)] $ 3.5 \leq M < 5.95$,
\item[3)] $ M \geq 5.95$
\end{itemize}
at $z=0$ \citep[selected as in][]{Sanchez_08},
\begin{itemize}
\item[1)]  $M < 2.9$,
\item[2)] $ 2.9\leq M < 4.65$,
\item[3)] $ M \geq 4.65$
\end{itemize}
at $z=0.5$ and 
\begin{itemize}
\item[1)]  $M < 2.6$,
\item[2)] $ 2.6 \leq M < 3.8$,
\item[3)] $ M \geq 3.8 $
\end{itemize}
at $z=1$. These limits were chosen in order to include about half of the total number of haloes in mass range 1 and the remaining equally divided between samples 2 and 3. The number of haloes ($N_{\rm h}$) and the shot noise term ($1/\bar n$), a scale independent poisson term arising from the discretization of the density field (see also Appendix \ref{ap:power_spectrum}), for each mass bin at each redshift are given in Table \ref{tab:bins}.

\subsection{Power spectrum computation and shot noise} \label{Pk_comp}

In order to compute the power spectra, the dark matter particles or haloes in the simulations were assigned to a grid of $1008^{3}$ cells using the Triangular Shaped Cloud (TSC) as Mass Assignment Scheme (MAS). We then computed the Fourier transform of the obtained density field by a fast Fourier transform (FFT) algorithm using the free software FFTW\footnote{http://www.fftw.org/} \citep[Fastest Fourier Transform in the West,][]{Frigo_05}. We then correct the amplitude of the Fourier modes for the effects of the MAS as in the first line of equation \eqref{eq:Jing_taylor_jeo}, spherically average them in shells of thickness $\Delta k= 2\pi/L=0.0047\,h\,\rm Mpc^{-1}$ and subtract the shot noise contribution $1/\tilde{n}$. For this configuration the Nyquist wavenumber is $k_{\rm N}=2.36\,h\,\rm Mpc^{-1}$ and the computed power spectrum is exact for $k<67\%k_{\rm N} = 1.58\,h\,\rm Mpc^{-1}$ (see appendix \ref{ap:power_spectrum} for more details). To obtain the redshift space power spectrum we computed the apparent position of the dark matter particle or haloes converting their velocities along a single axis of the simulation into a displacement in comoving coordinates.

We computed the real and redshift-space power spectra of the dark matter distribution ($P_{\rm DM}$) at redshift 0, 0.5 and 1, and the corresponding power spectra of the total halo catalogue and the three mass sub-samples, which we label as  $P_{\rm tot}$, $P_{\rm 11}$, $P_{\rm 22}$ and $P_{\rm 33}$ respectively. We also computed the cross power spectra for the three possible combinations of the mass bins at each redshift ($P_{\rm 12}$, $P_{\rm 13}$ and $P_{\rm 23}$).

\begin{figure}
 \includegraphics[width=85mm, keepaspectratio]{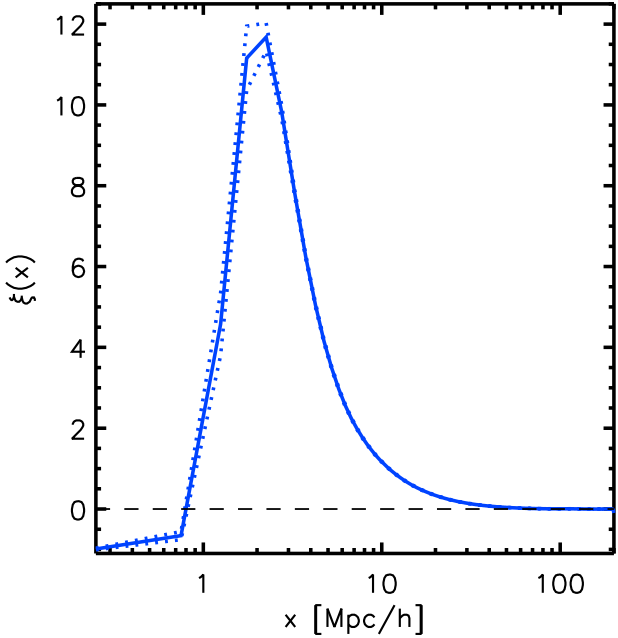}
 \caption{Log-linear scaling of the mean real-space correlation function $\xi(x)$ (solid line) of the total halo sample from the ensemble of simulations. The dotted lines indicate the variance from the different realisations. The signature of the exclusion effect in the halo sample can be clearly seen at $r < 2\,h^{-1}\,{\rm Mpc}$.} \label{fig:corr_func}
\end{figure}

\begin{figure}
 \includegraphics[width=85mm, keepaspectratio]{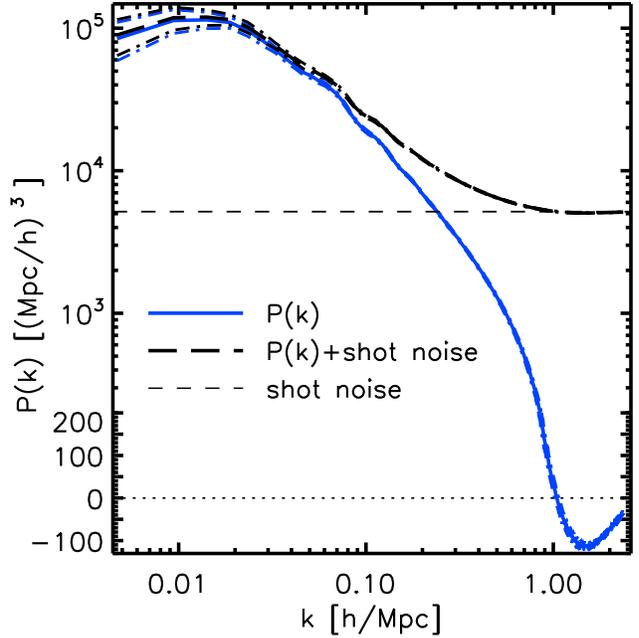}
 \caption{Mean power spectrum with and without the shot noise subtracted (solid and long-dashed lines, respectively) from the ensemble of simulations and their corresponding 1-$\sigma$ variance (dot-dashed lines) for the total halo sample. The amplitude of the shot noise is indicated by the horizontal dashed line. To enhance the negative part of the power spectrum at small scales, we use linear vertical axis for $P(k)<200\,h\,\rm Mpc^{-1}$ instead of logarithmic.}\label{fig:PS_excl}
\end{figure}

The FoF algorithm, used to create the halo catalogues, is intrinsically exclusive: two haloes must be separated by a distance larger that the sum of their radii or they would be identified as a single more massive halo. This introduces an exclusion effect in the halo catalogues which is visible in both the correlation functions and the power spectra of these samples. In the correlation function the exclusion effect is clearly visible at small
scales, where it becomes negative. As an example, Figure \ref{fig:corr_func} shows the mean correlation function obtained from the total halo catalogue at $z=0$ in our ensemble of simulations: after reaching
its maximum value at $r\approx2\,h^{-1}\,{\rm Mpc}$, it decreases converging to $\xi=-1$ as $r\rightarrow0$. Figure \ref{fig:PS_excl} shows the mean power spectrum with and without the shot noise subtracted (solid and long-dashed lines, respectively) for the same sample and redshift. The dot-dashed lines show their corresponding $1-\sigma$ variance. The shot noise amplitude is indicated by the horizontal dashed line. It is important to notice that when the shot noise is subtracted, the power spectrum becomes negative for $k\gtrsim 1\,h\,\rm Mpc^{-1}$. We have tested that this feature is independent on the shot noise amplitude or on the dimension of the grid used for the FFT. This clearly points to the fact that in the presence of the exclusion effect the noise is no more Poisson and possibly scale-dependent.

The problem of the existence and impact of the exclusion effect in the power spectrum has been already addressed in previous analyses. \citet{Smith_07} point out that the exclusion effect may give rise to a scale dependent noise term which may lead to a misinterpretation of the shape of the power spectrum; \citet{casas_02} and \citet{manera_09} found that biased objects present non Poisson noise. Since the understanding and modelling of the influence of the exclusion effect on the power spectrum is beyond the scope of this paper, we will not address specifically this problem here. Nonetheless, as will be explained section \ref{model}, our model of the shape of the power spectrum contains few free parameters that can partially absorb possible deviations from white noise. We will come back on this issue in section \ref{conclusion}.

\section{Modelling the full shape of $P(k)$}\label{model}

\subsection{The model}
\label{sec:model}

The increasing volume of the new large galaxy redshift surveys requires accurate models of the LSS observations in order to extract the maximum amount of information from the data without introducing systematic effects. The two standard approaches used to model the power spectrum and the correlation function are the halo model \citep[see][for a review]{cooray_HM}, and the use of perturbation theory techniques \citep[see] [for a review]{Bernardeau_02}. At redshift $z>1$, perturbation theory is able to provide a correct description of the power spectrum of the dark matter clustering at mildly non linear scales when terms of at least third order in the density fluctuations are included \citep{Jeong_06,Jeong_09}. However, none of these approaches is enough to model the full shape of the power spectrum in the non-linear regime or at smaller redshifts.

In recent years, there has been substantial progress regarding the theoretical understanding of non-linear evolution using perturbation theory. Have been proposed several methods which are based on a partial resummation of high order terms both in Eulerian space as renormalized perturbation theory \citep[RPT,][]{crocce_RPT1, crocce_RPT2, bernardeau_08} or in Lagrangian space \citep[Lagrangian Perturbation theory: LPT,][]{mats_LPT1,mats_LPT2}. Similar approaches are based on the renormalization group equations \citep{matarrese_rengroup1,matarrese_rengroup2} or on the renormalization of the parameters of the model in order match observables \citep{mcdonald_renorm,mcdonald_renormbias,smith09_SZ}. A different approach to PT is based on breaking the infinite hierarchy of $n$-point statistics by introducing physically motivated closure equations at the desired order \citep{pietroni_flowtime, taruya_closure, taruya_09}.

Within the theoretical framework of RPT, the non-linear power spectrum of the density fluctuations can be computed as the sum of two terms containing different physical information.The first of these contributions contains all the terms in the perturbation theory expansion of $P(k,z)$ that are proportional to the initial linear theory power spectrum $P_{\rm L}(k)$ in the so called {\it renormalized propagator} $G(k, z)$, which represents a non-linear correction to the growth factor. The second contribution to $P(k, z)$ groups all the remaining terms into the \emph{mode coupling power spectrum} $P_{\rm MC}(k,z)$, that contains the power arising from the coupling of different Fourier modes. The resulting power spectrum can be written as 
\citep{crocce_RPT1, crocce_RPT2}
\begin{equation}\label{eq:rpt}
 P(k, z) = G(k,z)^2P_{\rm L}(k,z) + P_{\rm MC}(k,z),
\end{equation}
where the redshift evolution of $P_{\rm L}(k,z)$ is given by the linear theory growth factor $D(z)$.

In the large-$k$ limit the propagator is accurately described by a Gaussian damping
\begin{equation}\label{eq:propagator}
 G(k,z)=\exp\left[-\left(\frac{k}{\sqrt{2}k_{\star}} \right)^{2}\right],
\end{equation}
where the damping scale $k_{\star}$ is given by
\begin{equation}\label{eq:k_star}
 k_{\star}(z)=\left(\frac{1}{6\pi^{2}}\int {\rm d}k\,P_{\rm L}(k, z)\right)^{-1/2}.
\end{equation}
Equation \eqref{eq:propagator} also provides a good aproximation for the propagator for small values of $k$.

The mode coupling power spectrum is given by the sum of an infinite number of terms $P_{n\,\rm MC}(k)$ ordered by the number $n$ of initial modes coupled. The contribution of all the terms combining $n$ modes becomes important at progressively smaller scales as $n$ increases. The first term is thus $P_{2\,\rm MC}(k)$, the power generated by the coupling of two modes, which can be approximated by the standard PT term
\begin{equation}\label{eq:rpt_1loop_app}
 P_{\rm 1loop}(k) = \frac{1}{4\pi^{3}} \int {\rm d}^{3}q |F_{\rm2}(\mathbf k-\mathbf q,\mathbf q)|^{2}
P_{\rm L}(|\mathbf k-\mathbf q|)P_{\rm L}(q),
\end{equation}
where $F_{2}(\mathbf k,\mathbf q)$ is the second order kernel of perturbation theory \citep[eq. 45 in][]{Bernardeau_02}. In the 
literature the power in equation \eqref{eq:rpt_1loop_app} is usually reffered to as $P_{22}(k)$.

\begin{figure}
 \includegraphics[width=80mm, keepaspectratio]{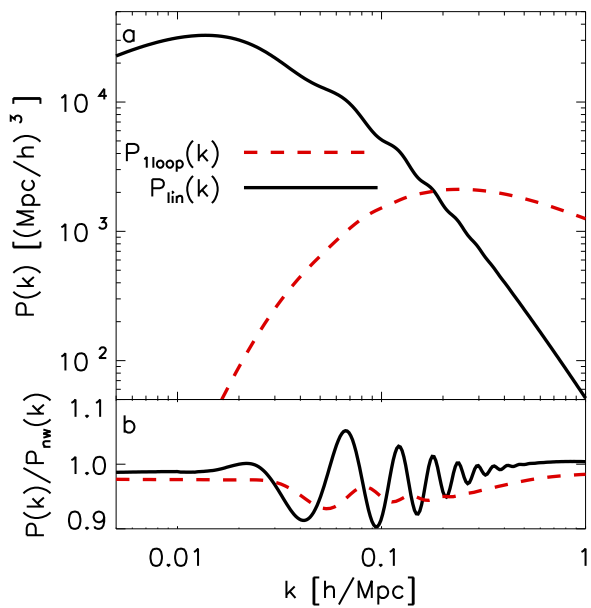}
 \caption{Panel a: Linear theory (solid line) and first loop (dashed line) power spectra. Panel b: ratio between the power spectra of panel a  and a reference power spectrum without oscillations. The first loop contribution $P_{\rm 1loop}(k)$ shows small oscillations out of phase with respect to $P_{\rm lin}(k)$, generating a net shift of the BAO peaks when summed.}\label{fig:PS_model}
\end{figure}

Panel \emph{a} of Figure \ref{fig:PS_model} shows the linear theory power spectrum (solid line) and $P_{\rm 1loop}(k)$ (dashed line) for the cosmological model of our ensemble of simulations. Panel \emph{b} shows $P_{\rm L}(k)$ divided by a reference power spectrum without BAO \citep{Eisenstein_98} and $P_{\rm 1loop}(k)$ divided by the corresponding term derived applying equation \eqref{eq:rpt_1loop_app} to the smooth power spectrum. $P_{\rm 1loop}(k)$ also contains oscillations, although with smaller amplitude and, more importantly, out of phase with respect to the ones in $P_{\rm L}(k)$. When these two terms are summed as in equation~\eqref{eq:rpt}, the BAO are shifted towards smaller scales with respect to the ones in the linear power spectrum \citep{crocce_nonlinBAO}. 

\citet{crocce_RPT2} showed that the predictions from RPT give an accurate description of the real space dark matter power spectrum measured from N-body simulations without the need to tune a free parameter. Despite this success, a drawback of this formalism is that there is no straightforward way to include the effects of redshift space distortions and bias, making its direct application to observational datasets impossible. However, \citet{crocce_nonlinBAO} proposed a model for the large scale correlation function motivated by the RPT formalism. In this ansatz the correlation function is given by
\begin{equation}
 \xi_{\rm NL}(r) = b^2 \left( \xi_{\rm L}(r)\otimes {\rm e}^{-(k_{\star}r)^2} 
+ A_{\rm MC} \,\xi'_{\rm L}(r)\,\xi^{(1)}_{\rm L}(r) \right), 
\label{eq:xi_model}
\end{equation}
where $b$, $k_{\star}$ and $A_{\rm MC}$ are treated as free parameters, and the symbol $\otimes$ denotes a convolution. Here $\xi'_{\rm L}$ is the derivative of the linear correlation function and $\xi^{(1)}_{\rm L}(r)$ is defined by 
\begin{equation}
 \xi_{\rm L}^{(1)}(r) \equiv \hat{r} \cdot \nabla^{-1}\xi_{\rm L}(r)
=\frac{1}{2\pi^2}\int P_{\rm L}(k)\,j_1(kr)k\,{\rm d}k ,
\label{eq:xi1}
\end{equation}
with $j_{\rm 1}(y)$ denoting the spherical Bessel function of order one. The second term in equation~\eqref{eq:xi1} corresponds to the leading order contribution to $\xi_{\rm MC}$ from the one-loop approximation of the mode coupling power of equation~\eqref{eq:rpt_1loop_app}. \citet{Sanchez_08} compared this model against the results of N-body simulations and found that it is able to give an accurate description of the full shape of the correlation function, including the effects of bias and redshift space distortions, for volumes 
up to two orders of magnitude larger than present day datasets. \citet{Sanchez_09} successfully used this model to obtain constraints on cosmological parameters from the correlation function of a sample of luminous red galaxies (LRG) drawn from the data release 6 (DR6) of the SDSS as measured by \citet{cabre_09}.

In our analysis, we follow the same approach and model the non-linear power spectrum as 
\begin{equation}\label{eq:rpt_mod}
 P(k,z) = b^{2} \left(e^{-\left(k/k_{\star} \right)^{2}} P_{\rm lin}(k,z) + A_{\rm MC}P_{\rm 1loop}(k,z)\right),
\end{equation}
and treat $b$, $k_{\star}$ and $A_{\rm MC}$ as free parameters. In the next section we will show that this model allows to obtain unbiased constraints on the dark energy equation of state parameter by accurately describing the full shape of the power spectrum measured in real and redshift space. Recently \citet{crocce_10} pointed out that, according to equation \eqref{eq:k_star}, $k_{\star}$ scales with the inverse of the growth factor $D(z)^{-1}$.

$P_{\rm 1loop}(k)$ does not have the same shape as $P_{2\,\rm MC}(k)$: the latter in fact decreases faster than the former and at $k\sim 0.15-0.2 \,h\,\rm Mpc^{-1}$ it is roughly $1.5-2$ times smaller. But at those scales the amplitude of the three mode coupling, $P_{3\,\rm MC}(k)$, is already around $1/4-1/2$ of $P_{2\,\rm MC}(k)$ and should be included in the model \citep[see Figure \ref{fig:PS_model} here and figure 1 in][for a comparison]{crocce_RPT1}. $P_{\rm 1loop}(k)$ is thus somewhat larger than $P_{2\,\rm MC}(k)+P_{3\,\rm MC}(k)$ and the difference becomes more and more important with increasing wave number. We will come back to this again in section \ref{p_k_fit}, where we will discuss the range of scales in which the model of equation \eqref{eq:rpt_mod} can be applied to a measurement of the power spectrum. 

We expect our model to be less efficient at describing the shape of the power spectrum in redshift-space than in real-space. In fact, we do not include redshift space distortion explicitly and we let the free parameters compensate some of its effects. Since the scale dependence of the redshift space distortions is stronger for the dark matter than for the haloes \citep{scoccimarro_04, Angulo_08}, the lack of a model for them will be particularly visible in the former case.

\section{Model in practice and discussion}\label{p_k_fit}

In this section we test whether or not the model described in section~\ref{sec:model} can give an accurate description of the power spectrum including the effects of non-linear evolution, redshift space distortions and bias. In section~\ref{sec:testing} we describe the test we implement to determine if the model returns unbiased constraints on the dark energy equation of state parameter $w_{\rm DE}$. Sections~\ref{pk}-\ref{sec:bias}
describe the results obtained when each of these scale dependent effects is included in the measurement of $P(k)$.

\begin{figure*}
\includegraphics[width=170mm, keepaspectratio]{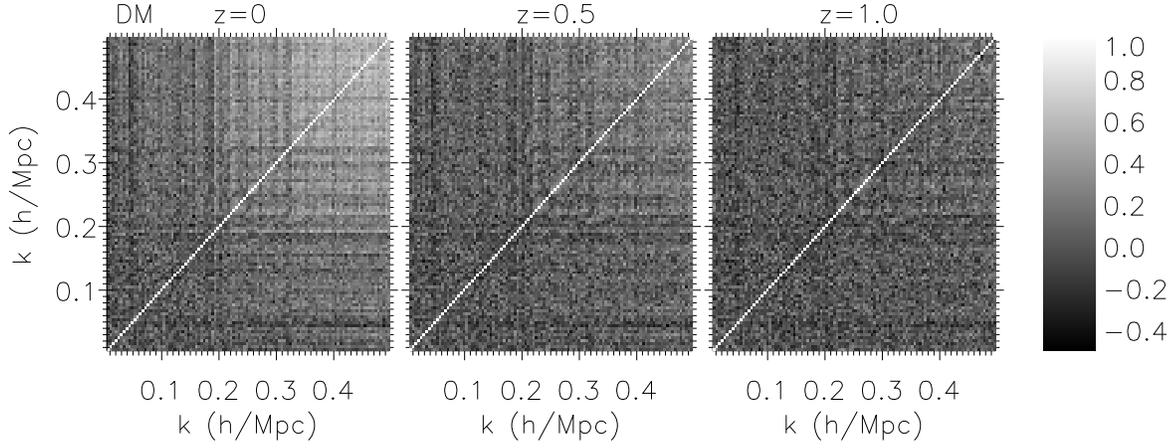}
\caption{Correlation matrix  $C_{ij}/\sqrt{C_{ii}C_{jj}}$, where $C_{ij}$ are the elements of the covariance matrix $\mathbfss C$, for the real-space dark matter power spectra at $z=0$ (left), $z=0.5$ (center) and $z=1$ (right). At $z=1$ the correlation matrix is close to diagonal, but at lower redshifts, when non-linearities become increasingly important, a significant correlation between different modes appears and becomes more important for larger scales, i.e. smaller $k$, as the redshift decreases. A similar result is obtained for the redshift-space power spectrum.}
\label{fig:PS_covariance_DM}
\end{figure*}
\begin{figure*}
\includegraphics[width=170mm, keepaspectratio]{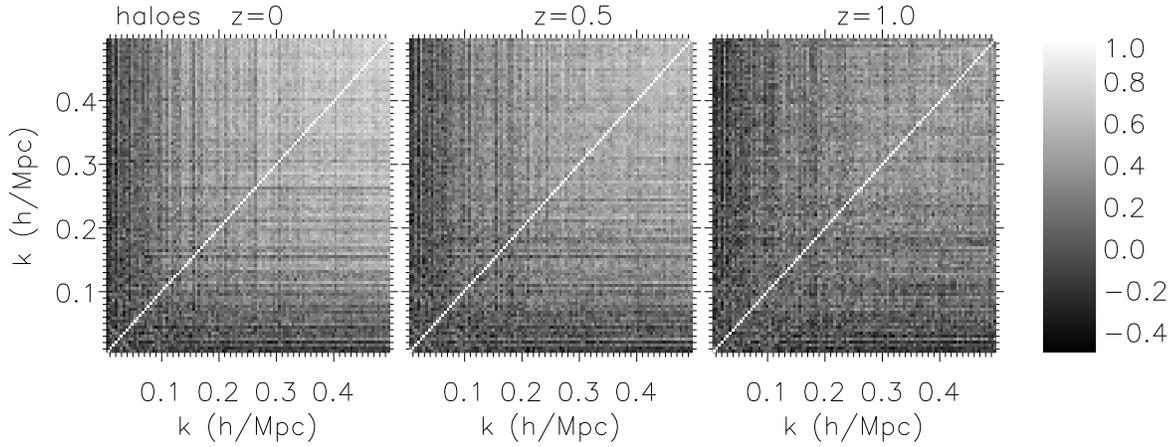}
\caption{Same as in Figure \ref{fig:PS_covariance_DM} but for the total halo samples at $z=0$ (left), $z=0.5$ (center) and $z=1$ (right) in real space. As for dark matter, the mode coupling becomes more important at larger scales as the redshift decreases. A similar result is obtained for the redshift space catalogue and for the other halo catalogues.} \label{fig:PS_covariance_hh}
\end{figure*}

\subsection{Testing the model}\label{sec:testing}

We now test whether or not the model for the power spectrum described in section~\ref{sec:model} returns
unbiased constraints on the dark energy equation of state parameter $w_{\rm DE}$. For this, we follow 
\citet{Angulo_08} and \citet{Sanchez_08} and consider the simple case in which we assume the values of
all other cosmological parameters to be known, and analyze the constraints on $w_{\rm DE}$ only.

When measuring the power spectrum from an observational dataset, it is necessary to assume a fiducial cosmology in order to map the observed galaxy redshifts and angular positions into comoving distances.
This choice has an impact on the results obtained. The use of a value of $w_{\rm DE}$ different from its true
value modifies the measured separations between the objects in the sample, changing the shape of the measured power spectrum. However, for small deviations away from the true equation of state, the alteration in the measured power spectrum can be represented by a rescaling of the wavenumber from $k_{\rm true}$ to
$k_{\rm app}$. This change can be encapsulated in a `stretch' factor $\alpha$, defined by 
\citep{Huff_07}
\begin{equation}
 \alpha=\frac{k_{\rm app}}{k_{\rm true}}.
\end{equation}

We analyse the constraints on the stretch parameter obtained from the mean power spectra of the L-BASICC II
simulations, which can be translated into constraints on $w_{\rm DE}$ as in \citet{Angulo_08}. All quoted allowed ranges for the constrained parameters correspond to the $68\%$ confidence level according to the variance from the ensemble of simulations.

We explore the parameter space defined by $\mathbf{\theta}=(k_{\star}, \, b_{\rm 1},\,A_{\rm{MC}}, \, \alpha )$ using a Markov chain Monte Carlo (MCMC) technique \citep{gilks_MCMC, christensen_00_MCMC}. We assume that the likelihood function follows a Gaussian form
\begin{equation}\label{eq:likely}
 \mathcal{L}\propto \exp\left(-\frac{1}{2}\chi^{2}({\mathbf \theta})\right),
\end{equation}
where
\begin{equation}
 \chi^{2}(\theta)=(\mathbf{d}-\mathbf{t}(\theta))^{T} {\textsf {\mathbfss C}}^{-1} (\mathbf{d}-\mathbf{t}(\mathbf{\theta}))
\end{equation}
is the standard $\chi^{2}$, in which $\mathbf{d}$ is an array containing the fitted data, in our case the computed power spectrum, $\mathbf{t}(\mathbf{\theta})$ contains the model computed for a given set of parameters $\theta$ and $\mathbfss{C}$ is the covariance matrix of the measurement.

We computed the covariance matrix ${\textsf {\mathbfss C}}$ for the dark matter and the halo power spectra from the L-BASICC II simulations as
\begin{equation}
 C_{ij}=\frac{1}{N_{\rm real}-1}\sum_{l=1}^{N_{\rm real}}(P_l(k_i)-\bar{P}(k_i))(P_l(k_j)-\bar{P}(k_j)),
\end{equation}
where $P_l(k_i)$ corresponds to the measurement of the power spectrum at the $i$-th $k$-bin in the $l$-th
realisation and $\bar{P}(k_i)$ corresponds to the mean power spectrum from the ensemble at the same wavenumber.

Figures \ref{fig:PS_covariance_DM} and \ref{fig:PS_covariance_hh} show the correlation matrices $C_{ij}/\sqrt{C_{ii}C_{jj}}$ of the power spectra of the real-space dark matter and total halo samples respectively.  In both figures, the three panels correspond, from left to right, to $z=0$, 0.5 and 1. In all cases we observe that the correlation between different modes, due to non linear mode coupling, is stronger at $z=0$ and decreases at increasing redshift. At $z=0$ the correlation matrix of the total halo sample shows strong correlations for $k \gtrsim 0.1 \,h\,\rm Mpc^{-1}$. In order to obtain robust constraints on $\alpha$ these correlations must be included when fitting our model to the results from the L-BASICC simulations. For a more detailed studies on the covariance of the power spectrum using theoretical model and large numerical simulations see \citet{hamilton_06, smith_09_corr, takahashi_09}.

On our parameter space we implemented flat priors given by
\begin{itemize}
 \item $0 \,h\,\rm{Mpc^{-1}}$$< k_{\star} < 0.35\,h\,\rm{Mpc^{-1}}$,
 \item $0\le A_{\rm{MC}} < 10$,
 \item $0.5 \le \alpha < 1.5$.
\end{itemize}
We analytically marginalize the bias parameter $b$ over an infinite flat prior using equation F2 in \citet{lewis_02}. In order to obtain an estimate of the amplitude of the model power spectrum, we compute a
value of $b$ and its variance by maximizing the likelihood function of equation~\eqref{eq:likely} while the other parameters are kept fixed to their mean values obtained from the MCMC.

\subsection{Non-linear evolution}\label{pk}

\begin{figure*}
\includegraphics[width=170mm, keepaspectratio]{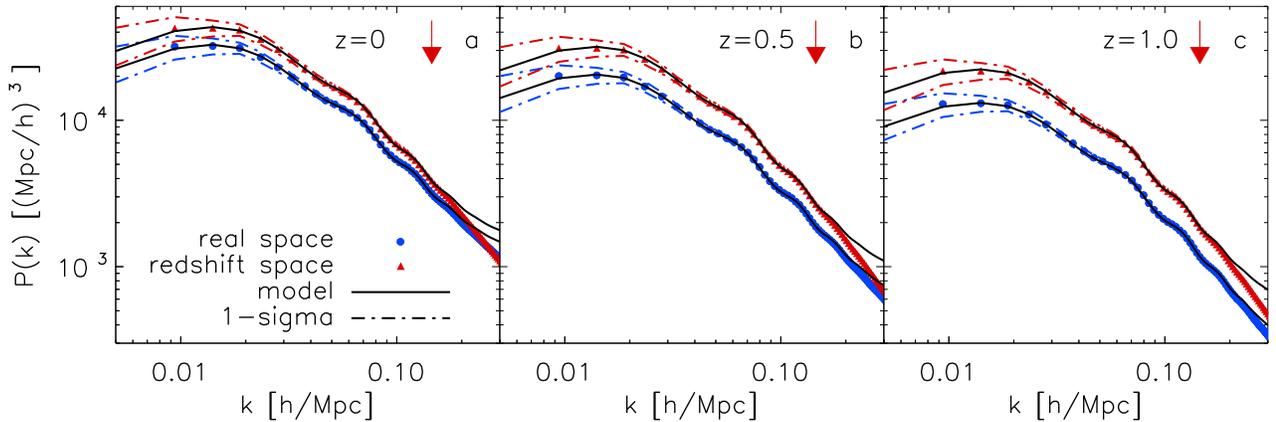}
 \caption{Mean power spectra computed from the simulations (circles for real space and triangles for redshift space), their variance (dash-dotted lines) and the model power spectrum as obtained through the fitting (solid lines) as function of the wavenumber in comoving units for the dark matter catalogue at redshift 0, 0.5 and 1.0 (left, centre and right respectively), in log-log scaling. The maximum wavenumber used for the fit is $k_{\rm max} = 0.15\,h\,\rm{Mpc^{-1}}$ and is indicated by the vertical arrow.}\label{fig:PS_computedDM}
\end{figure*}
\begin{figure*}
\includegraphics[width=170mm, keepaspectratio]{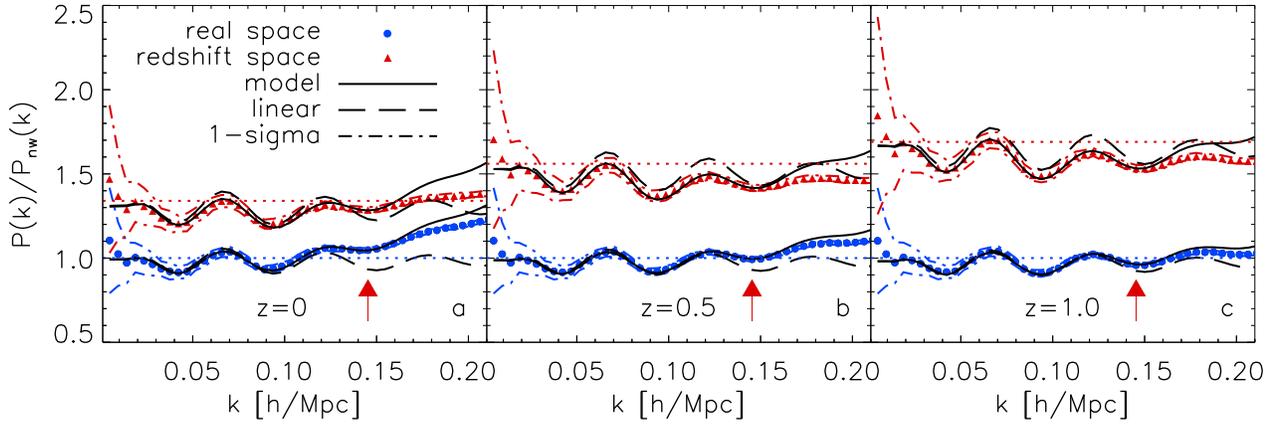}
 \caption{Power spectra of Figure \ref{fig:PS_computedDM} divided by a smooth reference power spectrum \citep{Eisenstein_98} in order to enhance the BAO oscillations, in linear scaling. For comparison the linear power spectrum, divided by the smooth power spectrum, is shown (dashed lines). The upper horizontal lines are drawn to the values corresponding to the Kaiser boost factor (section \ref{pk_red} and Table \ref{tab:bias}). The smooth power spectrum is the same for all the plots. The maximum wavenumber used for the fit is $k_{\rm max} = 0.15\,h\,\rm{Mpc^{-1}}$ and is indicated by the vertical arrow.}\label{fig:PS_computedDM_ratio}
\end{figure*}

\begin{figure*}
\includegraphics[width=170mm, keepaspectratio]{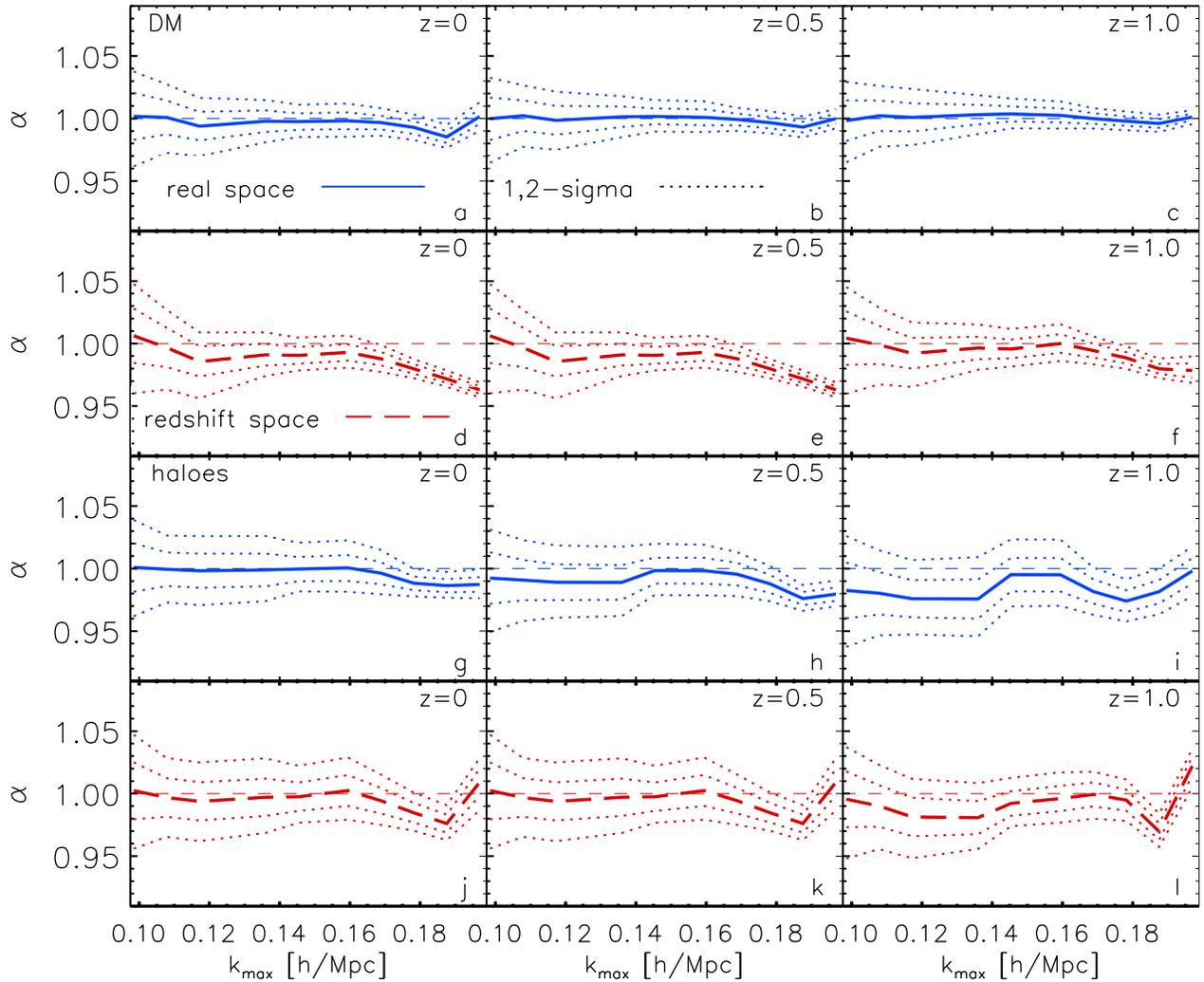}
 \caption{Mean value of $\alpha$ obtained from the MCMC (blue solid line for real space, red long-dashed line for redshift space), $1, 2-\sigma$ confidence level (dotted lines) as function of $k_{\rm max}$ for dark matter (real space: panels \emph{a}, \emph{b} and \emph{c}; redshift space: panels \emph{d}, \emph{e} and \emph{f}) and the total halo catalogue (real space: panels \emph{g}, \emph{h} and \emph{i}; redshift space: panels \emph{j}, \emph{k} and \emph{l}) at z=0, 0.5, 1 from left to right. The horizontal dashed  lines indicate $\alpha=1$.}\label{fig:errors_alpha}
\end{figure*}

\begin{figure*}
\includegraphics[width=170mm, keepaspectratio]{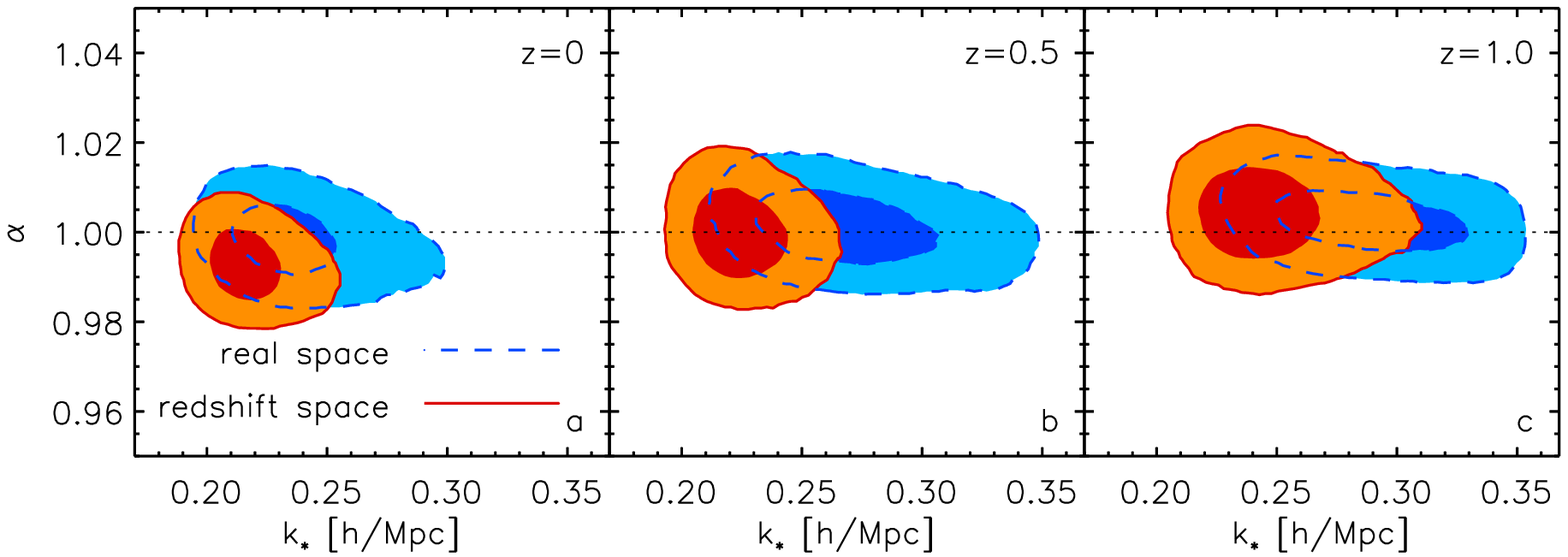}
 \caption{1 and 2 $\sigma$ contours of likelihood map in the $\alpha - k_{\star}$ plane obtained from the MCMC for dark matter at redshift 0 (left), 0.5 (centre), 1 (right), $k_{max}=0.15 \,h\,\rm{Mpc^{-1}}$. Background areas within dashed lines are for real space, foreground ones within solid lines are for redshift space.}\label{fig:maps_fitDM}
\end{figure*}

The circles in Figure~\ref{fig:PS_computedDM} show the mean real-space dark matter power spectra from the L-BASICC II simulations at $z=0$, 0.5 and 1 (panels \emph{a}, \emph{b} and \emph{c} respectively). Dot-dashed lines show the corresponding variances from the estimates in the different realizations. In order to highlight the signature of the acoustic oscillations, Figure~\ref{fig:PS_computedDM_ratio} shows, with the same symbols, the power spectra divided by a smooth linear theory power spectrum without BAO computed using the fitting formulas of \citet{Eisenstein_98}. The comparison with the linear power spectrum (dashed lines) shows that non linearities change the broad band shape of the power spectrum al small scales and damp the BAO feature. We explored the constraints on the stretch parameter obtained by applying the model described in section~\ref{sec:model} to these measurements.

The lines in the panels \emph{a}, \emph{b} and \emph{c} of Figure~\ref{fig:errors_alpha} show the mean value of $\alpha$ obtained from the measurements at $z=0$, 0.5 and 1 (solid lines), together with its correspondent $68\%$ and $95\%$ confidence levels (dot-dashed lines), as function of $k_{\rm max}$, the maximum value of $k$ included in the analysis. As smaller scales (larger values of $k_{\rm max}$) are taken into account, the width of the allowed region for $\alpha$ decreases due to the larger number of modes included in the fit. At $z=0$, the mean value of $\alpha$ remains consistent with 1 at a 1-$\sigma$ level for 
$k_{\rm max} \lesssim 0.16 \,h\,\rm{Mpc^{-1}}$, with $\alpha=0.999\pm0.007$. At higher redshifts the value of
$k_{\rm max}$ for which this holds increases, due to the smaller impact of non-linearities. The first row of the upper part of Table~\ref{tab:alpha} lists the obtained values of $\alpha$ for $k_{\rm max} = 0.15 \,h\,\rm{Mpc^{-1}}$. The solid lines in Figures \ref{fig:PS_computedDM} and \ref{fig:PS_computedDM_ratio} show the model power spectrum of equation~\eqref{eq:rpt_mod} computed using the mean values of the four parameters obtained for this range in $k$. This model is able to accurately describe the effects of non-linear evolution in both the broad-band shape of the power spectrum and the damping of the acoustic oscillations, up to the maximum value of $k$ included in the fit, which is indicated by a vertical arrow.

\begin{table}
 \centering
 \begin{minipage}{85mm}
   \caption{Mean values of $\alpha$ and their 1-$\sigma$ c.l. as recovered from the mean dark matter and halo samples power spectra of the three redshifts outputs of the L-BASICC II simulations for $k_{max}=0.15 \,h\,\rm{Mpc^{-1}}$. The upper part is for real space, the lower for redshift space. See Section \ref{Pk_comp} for the definition of the different samples listed in the second column.} \label{tab:alpha}
  \begin{tabular}{ c c c c c c}
     \hline 
    & & $z=0$ & $z=0.5$ & $z=1$\\
     \hline   
     \multirow{8}{*}{$\alpha_{real}$} & DM & $0.999\pm0.007$ & $1.001\pm0.006$ & $1.001\pm0.006$\\
    &tot & $1.001\pm0.011$ & $0.995\pm0.011$ & $0.994\pm0.014$\\
    &11 & $0.999\pm0.014$ & $1.006\pm0.014$ & $0.999\pm0.017$\\ 
    &22 & $1.006\pm0.017$ & $1.006\pm0.013$ & $1.005\pm0.023$\\
    &33 & $1.003\pm0.016$ & $0.997\pm0.012$ & $0.987\pm0.016$\\
    &12 & $1.002\pm0.013$ & $1.004\pm0.01$ & $0.987\pm0.015$\\
    &13 & $1.002\pm0.01$ & $0.991\pm0.012$ & $0.991\pm0.018$\\
    &23 & $0.995\pm0.013$ & $1.004\pm0.014$ & $1.005\pm0.013$\\
     \hline 
     \multirow{8}{*}{$\alpha_{red}$} & DM & $0.994\pm0.007$ & $1.001\pm0.007$ & $1.005\pm0.008$\\
    &tot & $1.002\pm0.012$ & $0.997\pm0.01$ & $0.987\pm0.012$\\
    &11 & $0.999\pm0.016$ & $1.008\pm0.016$ & $0.998\pm0.016$\\ 
    &22 & $1.007\pm0.02$ & $1.01\pm0.018$ & $1.006\pm0.026$\\
    &33 & $0.998\pm0.017$ & $0.994\pm0.014$ & $0.981\pm0.015$\\
    &12 & $1.002\pm0.014$ & $1.006\pm0.012$ & $0.982\pm0.014$\\
    &13 & $1.008\pm0.011$ & $0.993\pm0.012$ & $0.978\pm0.018$\\
    &23 & $0.997\pm0.015$ & $1.006\pm0.014$ & $1.006\pm0.014$\\
    \hline
  \end{tabular}
 \end{minipage}
\end{table}

\begin{figure*}
\includegraphics[width=170mm, keepaspectratio]{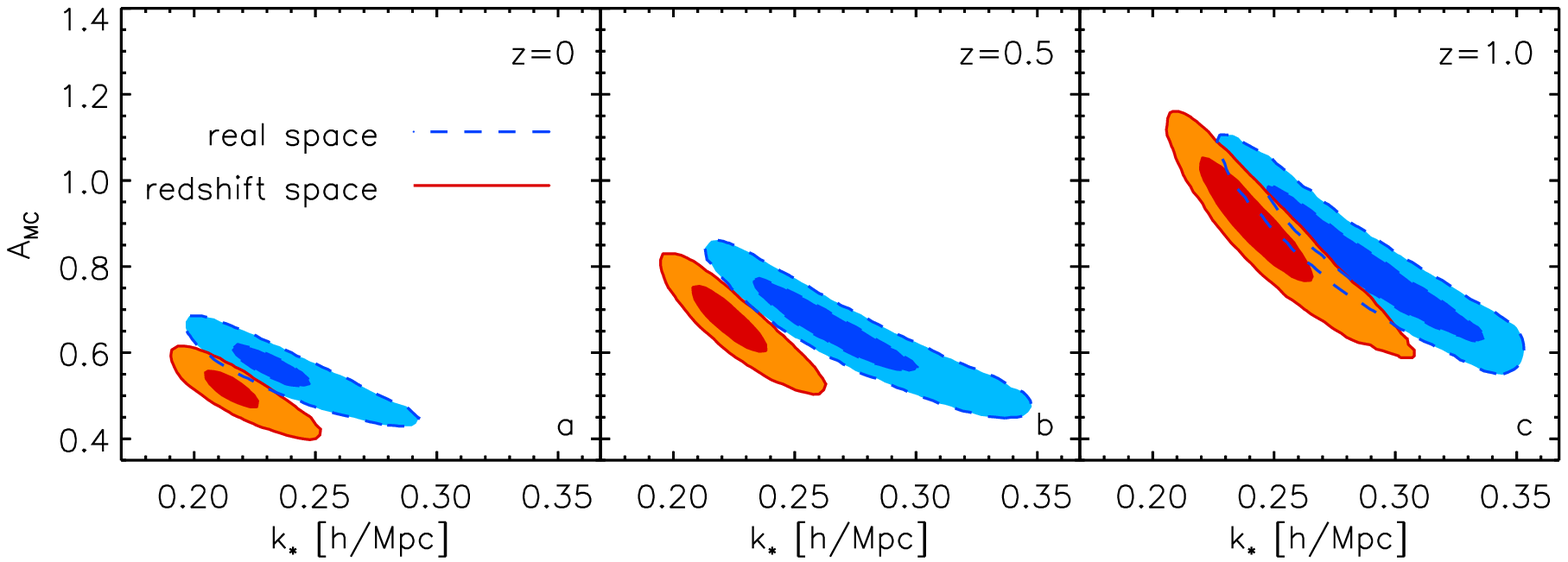}
 \caption{1 and 2 $\sigma$ contours of likelihood map in the $A_{MC} - k_{\star}$ plane obtained from the MCMC for dark matter at z=0 for $k_{max}=0.15 \,h\,\rm{Mpc^{-1}}$. Background areas within dashed lines are for real space, foreground ones within solid lines are for redshift space.}\label{fig:maps_kA_DM}
\end{figure*}

The dashed contours in Figure~\ref{fig:maps_fitDM} show the two-dimensional marginalized constraints in the $\alpha - k_{\star}$ plane obtained for $k_{\rm max} = 0.15 \,h\,\rm{Mpc^{-1}}$ at $z=0$, 0.5 and 1 (panels \emph{a}, \emph{b} and \emph{c} respectively). The contours correspond to $\Delta \chi^{2} = 2.3$ and $6.17$ which, assuming a two-dimentional gaussian likelihood, are equivalent to the 68\% and 95\% confidence levels.
While the constraints on the stretch parameter are very tight, there is a wide allowed region for $k_{\star}$ whose mean value is larger than the theoretical prediction of Equation~\eqref{eq:k_star}. The mean value $k_{\star}$ shows a tendency to increase with redshift, as the BAO feature is less damped, but the allowed range for this parameter is too large to compare this evolutions with $D(z)^{-1}$. The wide allowed range for $k_{\star}$ is caused by a strong degeneracy between this parameter and $A_{\rm{MC}}$. This can be seen in Figure~\ref{fig:maps_kA_DM}, which shows the two-dimensional constraints in the $A_{\rm{MC}}-k_{\star}$ plane for the dark matter power spectrum at $z=0$, 0.5 and 1. The degeneracy arises because it is possible to provide a good description of the overall shape of the power spectrum by compensating an increase in the damping of the first term of equation~\eqref{eq:rpt_mod} (a decrease of $k_{\star}$) by increasing the amplitude of the mode coupling contribution (using a higher value of $A_{\rm MC}$). Besides this, the value of $A_{\rm{MC}}$ obtained at $z=0$ is about $30\%$ smaller than the expected, confirming that the $P_{\rm{1loop}}(k,z)$ is somewhat bigger than the sum of the RPT two and three mode coupling. This makes the value of $k_{\star}$ slightly larger than the theoretical value of equation \eqref{eq:k_star}. In Figure~\ref{fig:maps_kA_DM} it is also noticeable that the values of $A_{\rm{MC}}$ increase with redshift due to the smaller relative amplitude of $P_{\rm 1loop}(k)$, which decreases as the growth factor squared,  with respect to the linear power spectrum.

\subsection{Redshift space distortions}\label{pk_red}

In galaxy surveys distances are inferred from the measured redshifts. The observed redshift of a galaxy is given by the sum of the contribution from the cosmological expansion and the Doppler shift due to its peculiar motion along the line of sight. This gives rise to a difference between the real and the apparent position of a galaxy, leading to an increase in the amplitude and a change of shape in the measured power spectrum $P_{\rm s}(k)$ with respect to its real-space counterpart.

On very large scales, peculiar velocities are dominated by the coherent flow of matter towards over-dense
regions. In linear theory, under the plane parallel approximation, this produces a scale-independent boost in
the amplitude of the large scale power spectrum given by \citep{kaiser_87}
\begin{equation} \label{eq:kaiser_theory}
 S_{\rm lin}=\frac{P_{\rm s}(k)}{P(k)}=1+\frac{2}{3}\beta+\frac{1}{5}\beta^{2},
\end{equation}
where $\beta=f/b$, $b$ is the bias factor and $f$ is the logarithmic derivative of the growth factor $D$ with respect to the scale factor $a$
\begin{equation}
 f=\frac{{\rm d}\ln D}{{\rm d}\ln a}.
\end{equation}

The linear theory description of equation~(\ref{eq:kaiser_theory}) is only valid asymptotically on extremely  large scales \citep{scoccimarro_04, Angulo_08, Jennings_2010}. On smaller scales, peculiar velocities are dominated by the random motions inside virialized structures which make bound haloes to appear elongated along the line of sight when mapped in redshift-space, an effect commonly known as ``fingers of god''. This causes a damping of the power spectrum that introduces deviations from the simple description of equation~(\ref{eq:kaiser_theory}) even at scales $k > 0.03\,h\,\rm{Mpc^{-1}}$. 

Figure~\ref{fig:PS_computedDM} shows the mean redshift-space dark matter power spectra (triangles) from the L-BASICC II simulations at $z=0$, 0.5 and 1 (panels \emph{a}, \emph{b} and \emph{c} respectively), together with their correspondent variances from our ensemble of simulations (dot-dashed lines). Figure~\ref{fig:PS_computedDM_ratio} shows the same power spectra divided by a smooth linear theory power spectrum without acoustic oscillations computed using the fitting formulas of \citet{Eisenstein_98}. Since the smooth power spectrum is the same both in real and redshift-space, it is possible to see both the increase in amplitude  and the change in shape towards smaller scales in the latter case. In this section we test if the parameterization of the model of equation~(\ref{eq:rpt_mod}) contains enough freedom to take into account these distortions. As stated in section~\ref{sec:model}, we can anticipate that the the performance of the model will be worse in this case than when dealing with real-space measurements, as we do not include explicitly the effect of redshift space distortions.

The dashed lines in the panels \emph{d}, \emph{e} and \emph{f} of Figure~\ref{fig:errors_alpha} show the mean value of the stretch parameter obtained by applying the model of equation~(\ref{eq:rpt_mod}) to the mean redshift-space power spectra as a function of $k_{\rm max}$ at $z=0$, 0.5 and 1, respectively. The dot-dashed lines indicate the correspondent $68\%$ and $95\%$ confidence levels. At $z=1$, also when dealing with redshift-space information the model is able to recover constraints on $\alpha$ consistent with one at the $1-\sigma$ level for $k < 0.18 \,h\,\rm{Mpc^{-1}}$. The constraints degrade at lower redshifts as non-linear redshift space distortions become more important and, at $z=0$, our results are only marginally consistent with $\alpha=1$. The first row of the lower part of Table~\ref{tab:alpha} lists the constraints on $\alpha$ obtained for $k_{\rm max} = 0.15 \,h\,\rm{Mpc^{-1}}$. In all cases, the allowed region for this parameter increases with respect to the real-space case. In all the panels of Figures~\ref{fig:PS_computedDM} and \ref{fig:PS_computedDM_ratio} the solid upper (orange) line shows the model power spectrum in redshift space with the parameters fixed to the mean value obtained from the MCMC.

\begin{table}
 \centering
 \begin{minipage}{85mm}
  \caption{Effective linear bias ($b_{\rm eff}$) computed from the halo model prescription, values of $b$ obtained from the fit in real space as described at the end of section~\ref{sec:testing}, theoretical linear Kaiser boost factors $S_{lin}$ (equation \ref{eq:kaiser_theory}) and Kaiser boost factors obtained from the fit ($S_{fit}$ equation \ref{eq:kaiser_ourmodel}) at all redshift, both for dark matter and the halo catalogues. All the ranges correspond to the $68\%$ confidence level.}  \label{tab:bias}
  \begin{tabular}{ c c c c c c }
  \hline
  & & $b_{\rm eff}$ & $b $ & $S_{\rm lin}$ & $S_{\rm fit}$\\
  \hline 
  \multirow{8}{*}{$z=0$}  & DM & 1 & $1.003\pm 0.003 $ & 1.34 & $1.32\pm0.01$ \\
    &tot & 1.89 & $1.89 \pm 0.01$ & 1.17 & $1.17\pm0.01$\\
    &11 & 1.63 & $1.65\pm0.01$ & 1.20 & $1.20\pm0.02$ \\ 
    &22 & 1.90 & $1.84\pm0.01$ & 1.17 & $1.17\pm0.02$ \\
    &33 & 2.54 & $2.49\pm0.01$ & 1.12 & $1.12\pm0.02$ \\
    &12 & 1.76 & $1.75\pm0.01$ & 1.18 & $1.17\pm0.02$ \\
    &13 & 2.04 & $2.04\pm0.01$ & 1.16 & $1.17\pm0.01$ \\
    &23 & 2.20 & $2.12\pm0.01$ & 1.14 & $1.15\pm0.02$ \\   
   \hline 
   \multirow{8}{*}{$z=0.5$}& DM & 1 & $0.999\pm 0.003$ & 1.56 & $1.55\pm0.01$\\
    &tot & 2.73 & $2.65\pm0.01$ & 1.18 & $1.18\pm0.01$ \\
    &11 & 2.31 & $2.33\pm0.01$ & 1.22 & $1.22\pm0.02$  \\ 
    &22 & 2.59 & $2.57\pm0.02$ & 1.19 & $1.19\pm0.02$ \\
    &33 & 3.42 & $3.39\pm0.02$ & 1.14 & $1.14\pm0.02$ \\
    &12 & 2.45 & $2.45 \pm 0.01$ & 1.20 & $1.21\pm0.02$ \\
    &13 & 2.81 & $2.79 \pm 0.01$ & 1.18 & $1.18\pm0.01$ \\
    &23 & 2.98 & $2.95 \pm 0.02$ & 1.17 & $1.17\pm0.02$ \\      
   \hline 
   \multirow{8}{*}{$z=1$} & DM  & 1 & $0.996 \pm 0.003$  & 1.69 & $1.70\pm0.01$ \\
    &tot & 3.87 & $3.80\pm0.02$ & 1.15 & $1.16\pm0.01$\\
    &11 & 3.27 & $3.37\pm0.02$ & 1.18 & $1.20\pm0.02$ \\ 
    &22 & 3.58 & $3.67\pm0.03$ & 1.17 & $1.18\pm0.03$ \\
    &33 & 4.62 & $4.73\pm0.03$ & 1.13 & $1.14\pm0.02$ \\
    &12 & 3.42 & $3.50\pm0.02$ & 1.17 & $1.18\pm0.02$ \\
    &13 & 3.89 & $3.98\pm0.02$ & 1.15 & $1.15\pm0.02$ \\
    &23 & 4.07 & $4.18\pm0.03$ & 1.14 & $1.15\pm0.02$ \\   
    \hline
  \end{tabular}
 \end{minipage}
\end{table}

As described in section~\ref{sec:testing}, we compute the value of $b$ and its variance fixing the other parameters to their mean values as obtained from the MCMC. Since this represents the large scale linear bias, we can calculate the Kaiser boost factor simply as:  
\begin{equation}\label{eq:kaiser_ourmodel}
 S_{\rm fit}=\frac{b^{2}_{\rm s}}{b^{2}}, 
\end{equation}
with $b$ and $b_{\rm s}$ the real and redshift-space linear bias. The values of $S_{\rm fit}$ obtained for the dark matter case are listed, together with the theoretical values $S_{\rm lin}$, in Table~\ref{tab:bias}. The three values of $S_{\rm lin}$ are also indicated in Figure \ref{fig:PS_computedDM_ratio} with the upper dotted lines. As can be seen the agreement between the theoretical and the recovered values are excellent at all the redshifts.

\begin{figure*}
\includegraphics[width=145mm, keepaspectratio]{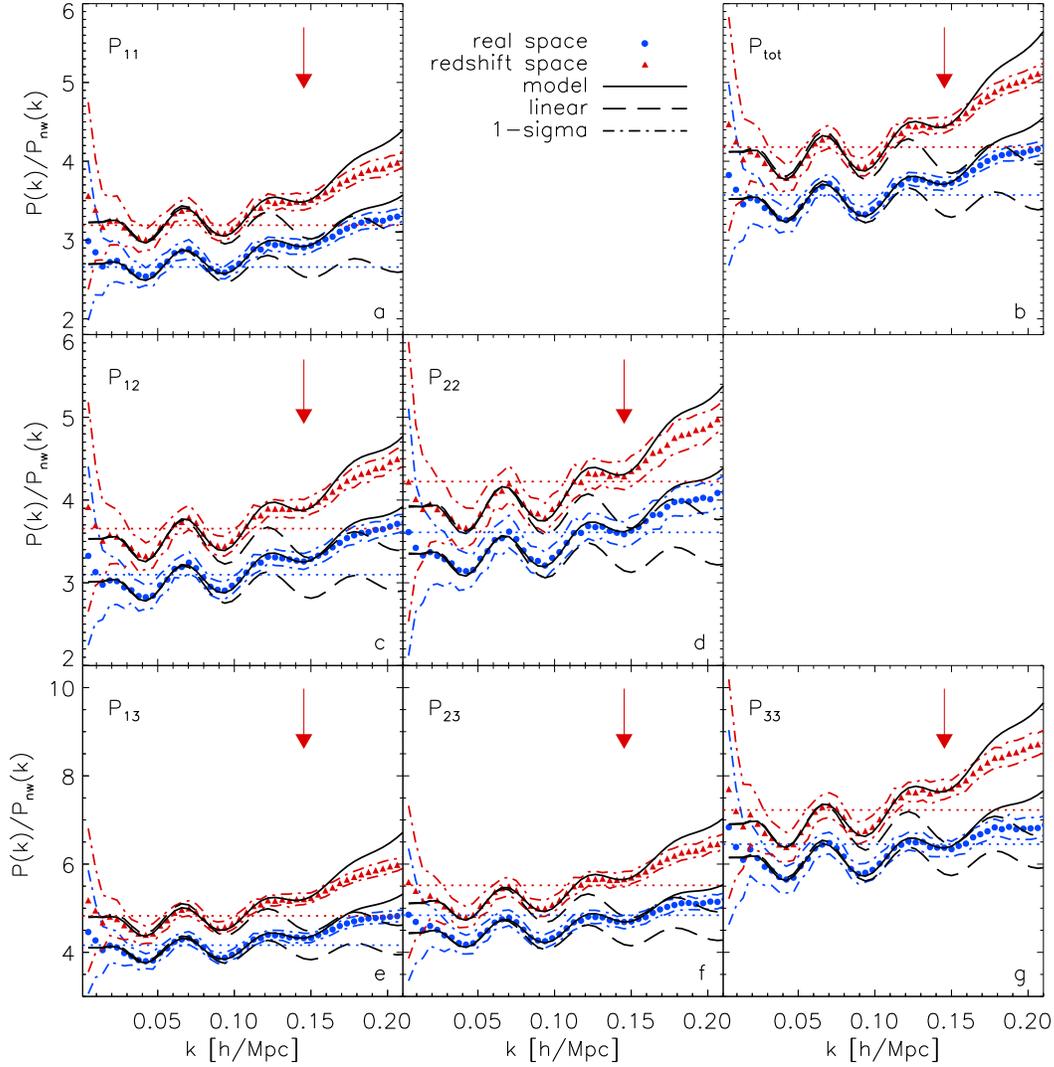}
 \caption{Mean power spectra computed from the simulations (circles for real space and triangles for redshift space), their variance (dash-dotted lines), the model power spectrum as obtained through the fitting (solid lines) and the linear power spectrum (dashed lines) as function of the wavenumber in comoving units at redshift 0 for the total catalog (\emph{tot}), the three mass bins (\emph{11}, \emph{22} and \emph{33}) and the three cross mass bins (\emph{12}, \emph{13} and \emph{23}) divided by a smooth reference power spectrum \citep{Eisenstein_98}. Lower horizontal lines indicate the linear bias, $b^{2}$ for the halo mass bin as computed from the halo catalogue prescription, the upper one indicate $b^{2}S_{\rm lin}$ (Table \ref{tab:bias}). The maximum wavenumber used for the fit is $k_{\rm max} = 0.15\,h\,\rm{Mpc^{-1}}$ and is indicated by the vertical arrow.}\label{fig:PS_computed0}
\end{figure*}

The shaded contours within solid lines in Figure~\ref{fig:maps_fitDM} show the two-dimensional 68\% and 95\% marginalized constraints in the $\alpha - k_{\star}$ plane obtained from the mean redshift-space power spectrum from the L-BASICC II simulations for $k_{\rm max} = 0.15 \,h\,\rm{Mpc^{-1}}$ at  $z=0$ (panel \emph{a}), 0,5 (\emph{b}) and 1 (\emph{c}). As in the case of real-space information, there is no degeneracy between these parameters. Redshift space distortions increase the damping of the BAO signal. This is reflected in the mean values of $k_{\star}$ being systematically lower than the ones obtained from real-space data. Figure~\ref{fig:maps_kA_DM} shows the correspondent constraints in the $k_{\star}-A_{\rm{MC}}$ plane. The scale-dependent effects introduced by redshift-space distortions cause these two parameters to follow a different degeneracy than in the real-space case although the qualitative behavior is maintained.


\subsection{Halo bias}\label{sec:bias}

\begin{figure*}
\includegraphics[width=145mm, keepaspectratio]{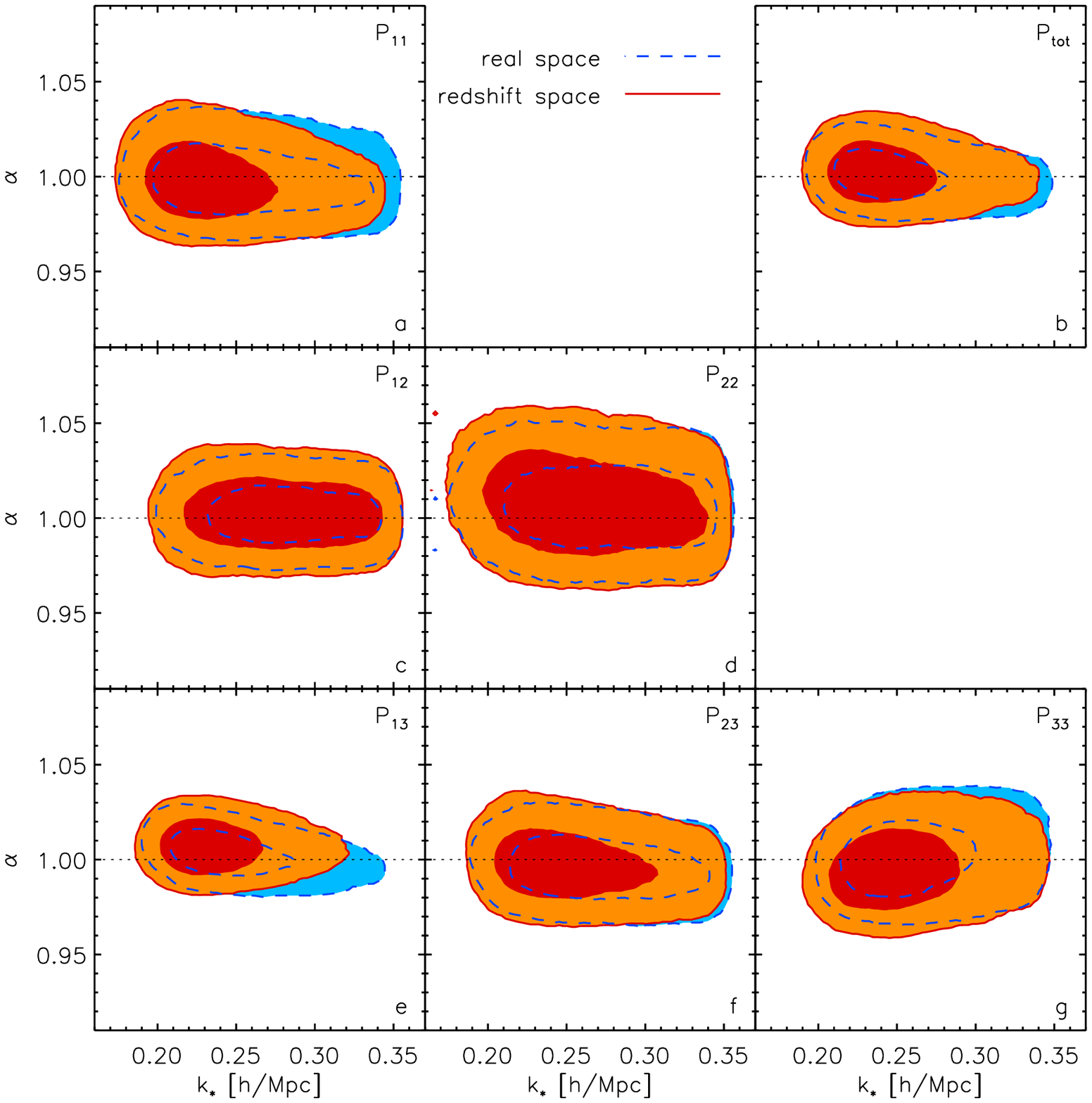}
 \caption{1 and 2 $\sigma$ contours of the likelihood map obtained from the MCMC for the parameters $\alpha$ and $k_{\star}$ for all the haloes catalogues at redshift 0 for $k_{\rm max}=0.15 \,h\,\rm{Mpc^{-1}}$. Background areas within dashed lines are for real space, foreground ones within solid lines are for redshift space.}\label{fig:maps_fit0}
\end{figure*}


In previous sections we compared the model of equation~(\ref{eq:rpt_mod}) against the clustering of the dark matter distribution. However, when dealing with real observational data sets, only the clustering pattern of the galaxy distribution can be used to obtain cosmological information. The relationship between the galaxy and matter density fields can be highly non-trivial, adding an extra complication to the interpretation of clustering measurements in terms of constraints on cosmological parameters \citep[e.g. see][]{sanchez_cole}.

A general galaxy sample is composed of central and satellite galaxies. The former are assumed to lie at or near the centre of the dark matter haloes while the later can be associated with the sub-structure present in more massive haloes that already contain a central galaxy. In redshift space, the signature of the fingers-of-god effect can be associated with the contribution of the satellite galaxies, which then are responsible for the most important non-linear redshift space distortions. \citet{tegmark04} analysed the SDSS DR2 compressing the fingers-of-god into isotropic structures. Recently \citet{reid_09} proposed a method to reconstruct the underlying halo density field from a galaxy sample by identifying fingers-of-god like structures and replacing them by one single halo. This procedure effectively eliminates the contribution of the satellite galaxies, which would correspond to the one-halo term in the halo-model formalism, from the measured power spectrum. This minimizes the impact of non-linear redshift-space distortions in the shape of the measured $P(k)$. Hence, in order to extract cosmological information from these measurements, a model of the full shape of the halo power spectrum is required. \citet{reid_09} also proposed a model, calibrated against numerical simulations, that relates the halo power spectrum to the linear theory prediction for the dark matter distribution. In this section we compare the model of equation~(\ref{eq:rpt_mod}) against the halo power spectra of the L-BASICC II ensemble of simulations for different halo samples.

According to linear theory, a simple relation is expected to hold between the dark matter and halo power spectra characterized by a scale-independent bias factor $b_1$
\begin{equation}
P_{\rm hh}(k)=b_1^2\,P_{\rm DM}(k).
\end{equation}
The results of numerical simulations have shown that this simple picture is only valid on very large scales and that, due to the effects of non-linear evolution, halo bias is a strong function of both scale and halo mass \citep{Smith_07,Angulo_08}. These distortions must be carefully modelled in order to obtain unbiased constraints on cosmological parameters from the full shape of the halo power spectrum. 

Figure~\ref{fig:PS_computed0} shows the mean power spectra at $z=0$ of the total halo sample of the simulations, $P_{\rm tot}(k)$ (panel \emph{b}), the power spectra of the three mass bins described in section~\ref{L-Basicc}, $P_{\rm 11}(k)$, $P_{\rm 22}(k)$ and $P_{\rm 33}(k)$ (panels \emph{a}, \emph{d} and \emph{g}), and their respective cross power spectra, $P_{\rm 12}(k)$, $P_{\rm 13}(k)$ and $P_{\rm 23}(k)$ (panels \emph{c}, \emph{e} and \emph{f}). To increase the dynamical range of the plot, these power spectra have been divided by the same non-wiggle linear-theory power spectrum as in Figure~\ref{fig:PS_computedDM_ratio}. Circles correspond to the real-space measurements while their redshift-space counterparts are shown by triangles, with the dot-dashed lines representing their respective variances from the ensemble of simulations.

We obtained constraints on the parameter space defined in section~\ref{sec:testing} for different values of $k_{\rm max}$ by applying the model of equation~(\ref{eq:rpt_mod}) to our measurements of the halo auto and cross power spectra. As example, Figure~\ref{fig:errors_alpha} shows the constraints on $\alpha$ as a function of $k_{\rm max}$ obtained from $P_{\rm tot}(k)$ in real (panels \emph{g}, \emph{h} and \emph{i}) and redshift-space (panels \emph{j}, \emph{k} and \emph{l}) for $z=0$, 0.5 and 1, respectively. The dot-dashed lines correspond to their 68\% and 95\% confidence levels. Due to the lower number densities of the halo samples, the variances of these power spectra are larger than for the dark matter case and increase with increasing redshift. This leads to an increase in the allowed regions for $\alpha$ with respect to the ones obtained using $P_{\rm DM}(k)$. Despite this difference, these results show a similar qualitative behavior to the constraints obtained from the dark matter power spectrum, with a mean value consistent with $\alpha=1$ up to $k \lesssim 0.15\,h\,\rm{Mpc^{-1}}$. The constraints on $\alpha$ obtained from the fit at different redshifts for $k_{\rm max}=0.15\,h\,\rm{Mpc^{-1}}$ are listed in Table~\ref{tab:alpha}. The solid lines in Figure \ref{fig:PS_computed0} show the model power spectrum of equation~\eqref{eq:rpt_mod} computed using the mean values of the four parameters obtained for this value of $k_{\rm max}$. 

The absence of a non-linear contribution from the fingers-of-god effect makes the scale dependence of redshift space distortions much less significant for the halo power spectrum than for the dark matter case \citep{Smith_07, Angulo_08, mats_LPT2}. This can be clearly seen in Figure~\ref{fig:PS_computed0}, where the differences in the shape of the real and redshift-space power spectra are small. This is reflected on the constraints on the stretch parameters, which in the redshift-space case show a similar behavior to those obtained using the real-space measurements.

Figure~\ref{fig:maps_fit0} shows the two-dimensional constraints in the $\alpha-k_{\star}$ plane obtained using the mean auto and cross power spectra for the different halo samples at $z=0$. The dashed lines show the 68\% and 95\% confidence limits obtained from real-space information while the solid lines correspond to the redshift-space case. The difference between the constraints obtained in the two cases is less significant than in the dark matter case shown in Figure~\ref{fig:maps_fitDM}. The constraints on $k_{\star}$ are broad due to the degeneracy between this parameter and $A_{\rm MC}$; nonetheless $k_{\star}$ seems to prefer smaller values in redshift space, in accordance with the more important damping of the BAO features.

In contrast to the behavior found in the analysis of the dark matter power spectrum, when applied to the different halo samples the ability of the model to obtain unbiased constraints on the stretch parameter degrades with increasing redshift and at $z=1$ they are only marginally consistent with $\alpha=1$. This is related to the mass resolution of the L-BASICC II simulations: the smallest halo mass that can be resolved corresponds to $M_{\rm halo} = 1.75 \times 10^{13} h^{-1} \rm M_{\odot}$. While at $z=0$ these haloes are moderately biased tracers of the underlying dark matter distribution, at $z=1$ they are much rarer objects that reside in very dense,
and thus highly non-linear, regions. Hence, an accurate description of the shape of the power spectrum of such high mass objects at high redshifts requires a more detailed model of the non-linear distortions
than that of equation~\eqref{eq:rpt_mod}.


The shape of the cross power spectra $P_{\rm12}(k)$, $P_{\rm13}(k)$ and $P_{\rm23}(k)$ can also be accurately described by the model of equation~(\ref{eq:rpt_mod}). As can be seen in Table~\ref{tab:alpha}, in all cases the constraints on $\alpha$ derived from these measurements are in good agreement with $\alpha=1$ and are tighter than the ones derived from their correspondent auto power spectra. 

As stated at in Section \ref{Pk_comp}, the halo catalogues are affected by the exclusion effect introduced by the Friend-of-Friends algorithm used to identify them. To account for a possible difference in the large scale amplitude of the shot-noise as could be caused by this effect, we have repeated the analysis adding a constant term to the model of equation \eqref{eq:rpt_mod} which we allowed to vary over a wide flat prior. Our findings indicate that such a term is degenerate with $k_{\star}$ and $A_{\rm MC}$, while the constraints of $\alpha$ are not affected.

We recover the values of the bias $b$ for all the power spectra in Figure~\ref{fig:PS_computed0} as indicated in section~\ref{sec:testing}. The real space biases are listed in Table~\ref{tab:bias} and $b^{2}$ is shown by lower dotted lines in Figure~\ref{fig:PS_computed0}. As a comparison we also list the theoretical values of the effective bias $b_{\rm eff}$ for the corresponding halo samples computed using the mass function proposed by \citet{jenkins_massfunc} and the halo bias prescription obtained from the spherical collapse model \citep{smt_2001}, which show a reasonable agreement with the measurements from the simulation with the exception of the high mass sample. Table~\ref{tab:bias} also shows the values of the Kaiser boost factor of equation~\eqref{eq:kaiser_theory} computed using $b_{\rm eff}$ and the ones obtained from the measured power spectra (equation \ref{eq:kaiser_ourmodel}): as for the dark matter case, they agree remarkably well. Since the dependency of equation~\eqref{eq:kaiser_theory} on $b$ is very weak, the difference between $S_{\rm lin}$ computed with $b_{\rm eff}$ and with $b$ is much smaller than the error bars of $S_{\rm fit}$. The upper dotted lines in Figure~\ref{fig:PS_computed0} correspond to the value of $b_{\rm eff}^{2}S_{\rm lin}$.

\section{Conclusions} \label{conclusion}

The increasing quantity and quality of information from large galaxy redshift surveys demands models able to describe the clustering of the galaxy distribution with high accuracy. The shape of the power spectrum, the tool most commonly used to analyze galaxy clustering, is distorted by non-linear evolution, redshift-space distortions and bias. These effects complicate the relation between the observations and the predictions of linear perturbation theory, making the interpretation of these measurements in terms of constraints on cosmological parameters more difficult.

In this paper we presented a model for the full shape of the large-scale power spectrum which is based on RPT and is analogous to the approach followed by \citet{crocce_nonlinBAO} and \citet{Sanchez_08} for the two-point correlation function. We compared our model against power spectra measured in a suite of 50 large volume, moderate resolution N-body simulations, called L-BASICC II \citep{Angulo_08, Sanchez_08}. Our results indicate that the simple model presented here can provide an accurate description of the full shape of the power spectrum taking into account the effects of non-linear evolution, redshift-space distortions and halo bias for scales $k\lesssim0.15 \,h\,\rm{Mpc^{-1}}$, making it a valuable tool for the analysis of real datasets.

When applied to the dark matter distribution, our model gives a better description of the shape of the power spectrum at increasing redshift, where non-linear effects become less important. Even though this holds both in real and redshift-space, in the latter case the model performs worse, since it does not include explicitly the effect of scale dependent redshift-space distortions. The model also gives a correct description of the shape of the halo power spectrum for different mass ranges both for real and redshift-space information. In the latter case, due to the absence of the non-linear contribution from the fingers-of-god effect, the scale dependence of the redshift-space distortions is simplified and the obtained constraints on the stretch parameter are similar to the ones of the real-space case.



\citet{Sanchez_08} performed a similar analysis for the shape of the large scale two-point correlation function. Their results showed that non-linear evolution, redshift-space distortions and bias are much more simpler to deal with for the case of the correlation function than for the power spectrum, where the signature is highly scale dependent. However, it is necessary to pursue complementary approaches to constrain cosmological parameters  allowing a comparison of the obtained results and a check for possible systematics. The model for the shape of power spectrum presented here has been proven to provide constraints on the stretch parameter $\alpha$ similar to the ones obtained by \citet{Sanchez_08}. Thus, with an accurate modelling, it is possible to extract cosmological information with the same precision from both statistics.

Using a similar set of simulations to the ones used in this work, \citet{Angulo_08} tested a model in which the information from the broad band shape of the power spectrum is discarded in order to extract a measurement of the BAO oscillations only. A comparison of their results with ours shows that the extra information in the full shape of the power spectrum helps to improve the obtained constraints over the BAO alone case \citep[see also][]{shoji_09}. In fact, despite the relatively small number of wave modes included in our analysis, the constraints that we obtain for $\alpha$ are slightly tighter than those of \citet{Angulo_08}. The model equation~(\ref{eq:rpt_mod}) can be extended in a number of ways, like for instance, with the use of the full RPT calculation of $P_{\rm2\, MC}$ and the inclusion of higher order terms in the description of $P_{\rm MC}$ or a specific treatment of the effects of redshift-space distortions. This can help to extend the range of scales in which it is valid and improve the accuracy of the constraints on $\alpha$ and thus the dark energy equation of state.

\section*{Acknowledgments}

We would like to give special thanks to Raul Angulo and Carlton Baugh for providing us with the L-BASICC II numerical simulations. We aknowledge Carlton Baugh and Martin Crocce for their comments on the draft. We thank Raul Angulo, Martin Crocce and Eiichiro Komatsu for useful discussions. FM thanks Lodovico Coccato, Maximilian Fabricius and Ralf K\"ohler for the technical help. FM was supported through SFB-Transregio 33 ``The Dark Universe'' by the Deutsche Forschungsgemeinschaft (DFG).

\appendix

\section{Power spectrum: FFT and corrections}\label{ap:power_spectrum}

The power spectrum is a statistical tool that allows to characterize gravitational clustering. In terms of the Fourier transform of the overdensity $\delta(\mathbf{x}, t)$
of the continuous density field $\rho(\mathbf{x}, t)$, its definition is
\begin{equation}
  (2\pi)^{3} \delta_{\rm D}^{3}(\mathbf{k}-\tilde{\mathbf{k}}) P(\mathbf{k}, z) = \langle \delta(\mathbf{k}, z)\delta^{*}(\tilde{\mathbf{k}}, z) \rangle,
\end{equation}
where $\delta_{\rm D}^{3}(\mathbf k)$ is the three dimensional Dirac delta
function. For a finite number $N$ of points tracing the underlying density field
in a limited volume $V$, the power spectrum becomes \citep{Jing_05}:	
\begin{equation} \label{eq:Jing_ft}
  P(k) = \langle |\delta(\mathbf{k})|^{2} \rangle - \frac{1}{\bar{n}},
\end{equation}
with $\bar{n}$ the mean number density of particles. The term $1/\bar{n}$ is commonly called shot noise, is white, i.e. scale independent, and is given by the discretization of the density field.

\begin{figure*}
  \includegraphics[width=175mm, keepaspectratio]{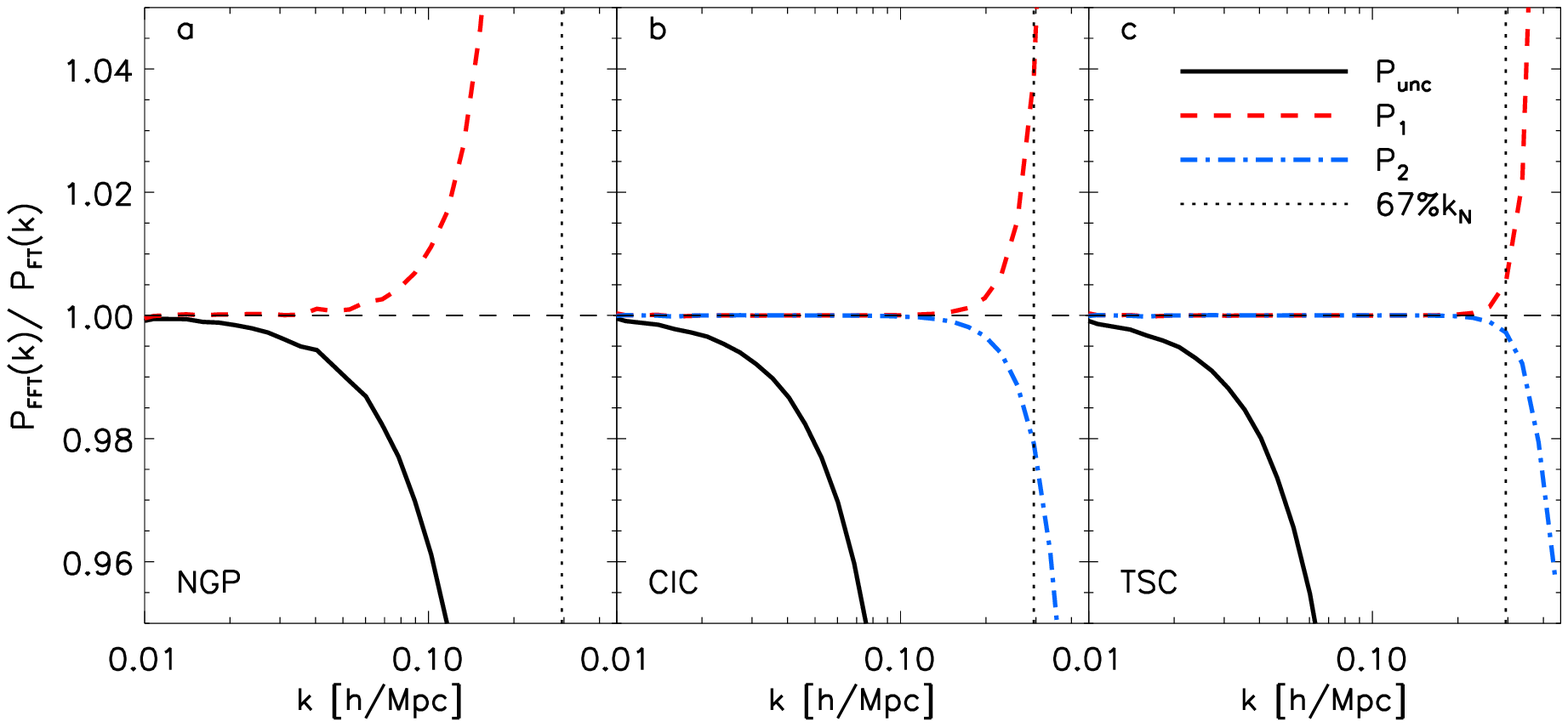}
  \caption{Ratio between the power spectrum computed using the FFT and the one computed with the standard FT versus the wavenumber $k$. For clearness the error bars are not shown (for $k\gtrsim0.03\,h\,\rm Mpc^{-1}$ their amplitude is comparable with the thickness of the lines). The FFT has been computed on a grid of $200^{3}$ cells, that has been filled using the NGP (left panel), CIC (central panel) and TSC (right panel). The solid black line is the ratio between the uncorrected $P_{\rm FFT}(k)$ and the $P_{\rm FT}(k)$, while the dashed red line and the dash-dotted blue line indicate the $P_{\rm FFT}(k)$ corrected as in equations \eqref{eq:Jing_taylor_ang} and \eqref{eq:Jing_taylor_jeo} respectively. Since $\sum_{\mathbf n} |W(\mathbf k + 2k_{\rm N}\mathbf n)|^{2}=1$, $P_{2}(k)$ is equivalent to the uncorrected power spectrum in the NGP case. The dotted vertical line outlines the $67\%k_{\rm N}$.} \label{fig:testMAS}
\end{figure*}

We Fourier transform the density field using a fast Fourier transform (FFT) algorithm that drastically increases the speed of the computation of the Fourier transform with respect to the standard transform. The drawback of the FFT is that it requires to convolve the density field $\delta(\mathbf{x})$ with a regular grid. 
The conversion of $\delta(\mathbf{x})$ into a density field $\delta^{\rm d}(\mathbf{x})$ on a grid of cell size $H$ is governed by the Mass Assignment Scheme (MAS) function  $W(\mathbf x)$.
When the power spectrum is computed convolving the density with the MAS, equation \eqref{eq:Jing_ft} becomes \citep{Hockney_81}
\begin{equation} \label{eq:Jing_fft}
  \begin{split}
    \langle |\delta^{\rm d}(\mathbf{k})|^{2} \rangle = &\sum_{\mathbf n} |W(\mathbf k + 2k_{\rm N}\mathbf n)|^{2}P(\mathbf k + 2k_{\rm N}\mathbf n) \\
  &+\frac{1}{\bar{n}} \sum_{\mathbf n} |W(\mathbf k + 2k_{\rm N}\mathbf n)|^{2},
  \end{split}
\end{equation}
where $\mathbf n$ is a three dimensional integer vector and $k_{\rm N} = \pi/H$ is the Nyquist wavenumber and represent the larger value of $\mathbf k$ that is possible to achieve with the FFT.  $W(\mathbf{k})$ is the Fourier transform of the MAS.

Three MAS most commonly used are the Nearest Grid Point (NGP), Cloud in Cell (CIC) and Triangular Shape Cloud (TSC). In three dimensions, they are given by $W(\mathbf x)=\prod_{i=1}^{3}W(x_{i})$, where:
\begin{subequations} \label{eq:MAS}
\begin{equation}
  W_{\rm NGP}(x_{i}) = 
  \begin{cases}
    1 & |x_{i}| < 0.5,\\
    0 & |x_{i}| \geq 0.5 ,
  \end{cases} 
\end{equation}
\begin{equation}
  W_{\rm CIC}(x_{i}) = 
  \begin{cases}
    1-|x_{i}| & |x_{i}| < 1,\\
    0 & |x_{i}| \geq 1,
  \end{cases} 
\end{equation}
\begin{equation}
  W_{\rm TSC}(x_{i}) = 
  \begin{cases}
    0.75 - x_{i}^{2} & |x_{i}| < 0.5,\\
    \frac{\left(1.5-|x_{i}|\right)^{2}}{2} & 0.5 \leq |x_{i}| < 1.5,\\
    0 & |x_{i}| \geq 1.5.
  \end{cases} 
\end{equation}
The Fourier transform of the MAS is given by
\begin{equation}
W(\mathrm k) = \left[\prod_{i=1}^{3} \frac{\sin(\pi k_{\rm i}/2k_{\rm N})}{\pi k_{\rm i}/2k_{\rm N}}\right]^{p},
\end{equation}
\end{subequations}
with p the order of the MAS ($p=1$ for NGC, $p=2$ for CIC and $p=3$ for TSC).

The computation of the power spectrum with a FFT algorithm is divided roughly in the following steps :
\begin{enumerate}
\item creation of a density field assigning the particles to a grid using a
chosen MAS;
\item Fourier transform of the density field - we have performed this step using the free software FFTW;
\item spherical average of the density field in Fourier space to obtain the left hand side of equation\eqref{eq:Jing_fft};
\item solve the equation \eqref{eq:Jing_fft} to obtain $P(k)$.
\end{enumerate}

In \citet{Jing_05}, the last point is solved with an iterative procedure based on the assumption that the PS can be approximated by a power law at $k > k_{\rm N}$. This assumption is a viable approximation when computing the power spectrum for dark matter, but it is not applicable when dealing with the halo-halo power spectrum (see Figure \ref{fig:PS_excl} and \S \ref{Pk_comp}). It is also possible to approximate the solution dividing the computed $\langle \delta^{\rm d}(\mathbf{k}) \rangle$ by $W(\mathbf k)$ (e.g. \citet{Angulo_08}) or by $(\sum_{\mathbf n} |W(\mathbf k + 2k_{\rm N}\mathbf n)|^{2})^{1/2}$ (e.g \citet{Jeong_09}). So the last two steps of the computation become:
\begin{enumerate}
\item[(iii)] correction of the amplitude of each Fourier mode for the effect of the MASS;
\item[(iv)] spherical average of the density field in Fourier space to obtain the left hand side of equation\eqref{eq:Jing_fft}.
\end{enumerate}
The recovered power spectra for the two corrections described above are
\begin{subequations}\label{eq:corr_ps}
\begin{equation} \label{eq:Jing_taylor_ang}
  \begin{split}
    P_{1}(k) =&\frac{\langle |\delta^{\rm d}(\mathbf{k})|^{2} \rangle}{W^{2}(\mathbf k)} - \frac{1}{\bar n} = P(k) \\ 
  &+ \sum_{|\mathbf n|\neq 0}\left[\frac{|W(\mathbf k')|^{2}P(\mathbf k')}{W^{2}(\mathbf k)}+\frac{1}{\bar n}\frac{|W(\mathbf k')|^{2}}{W^{2}(\mathbf k)}\right],
  \end{split}
\end{equation}
and
\begin{equation} \label{eq:Jing_taylor_jeo}
  \begin{split}
    P_{2}(k) =&\frac{\langle |\delta^{\rm d}(\mathbf{k})|^{2} \rangle}{\sum_{\mathbf n} |W(\mathbf k + 2k_{\rm N}\mathbf n)|^{2}} -\frac{1}{\bar n} = P(k) \\ 
    &\times\left[1+\frac{\sum_{|\mathbf n|\neq 0}|W(\mathbf k')|^{2}P(\mathbf k')}{W^{2}(k)P(k)} \right.\\
    &\left. - \frac{\sum_{|\mathbf n|\neq 0}|W(\mathbf k')|^{2}}{W^{2}(\mathbf k)}+ \ldots\right],
  \end{split}
\end{equation}
\end{subequations}
where $\mathbf{k}'=\mathbf{k} + 2k_{\rm N}\mathbf n$.
Equation \eqref{eq:Jing_taylor_jeo} makes use of the fact that $\sum_{\mathbf n} |W(\mathbf k')|^{2}$ is finite and that for $|\mathbf k| \lesssim k_{\rm N}$, $\sum_{|\mathbf n|\neq 0}|W(\mathbf k')| \leqslant |W(\mathbf k)|$ and $P(\mathbf k') \leqslant P(\mathbf k)$.

We have tested the impact of the MAS and corrections on the recovered power spectrum. To do this we have compared the power spectrum computed with the standard Fourier transform (FT) and with the FFTW algorithm. For the FFT we used a grid of $200^{3}$ cells, filled using one of the three possible MAS. We have also switched on and off the corrections in equations \eqref{eq:corr_ps}. In order to have the same number of modes in the two cases, the values of $\mathbf k$ used for the FT calculation are the same as for the FFT. The power spectrum for all the cases has been computed in 48 logarithmic bins between $k=0.01 \,h\,\rm Mpc^{-1}$ and $k_{\rm N}=0.47\,h\,\rm Mpc^{-1}$. Figure \ref{fig:testMAS} shows the results of this test as the ratio between the FFT and the FT power spectra as function of the wave-number $k$. The panels \emph{a}, \emph{b} and \emph{c} correspond, respectively, to the power spectrum computed with NGP, CIC and TSC as MAS. In each panel the solid line is for the uncorrected power spectrum $P_{\rm unc}(k) = \langle |\delta^{\rm d}(k)|^{2}\rangle - 1/\bar{n}$, the dashed line is for the $P_{1}(k)$ while the dash-dotted one is for $P_{2}(k)$. Since $\sum_{\mathbf n} |W(\mathbf k + 2k_{\rm N}\mathbf n)|^{2}=1$ in the NGP case, $P_{2}(k) = P_{\rm unc}(k)$. The correction of the $\langle |\delta^{\rm d}(k)|^{2}\rangle$ increases drastically the agreement with the real $P(k)$, and the achieved accuracy is larger the higher the order of the MAS is; equation \eqref{eq:Jing_taylor_jeo} seems to work slightly better than equation \eqref{eq:Jing_taylor_ang}. In the light of these results, we have decided to use the TSC as MAS correcting the power spectrum as in equation \eqref{eq:Jing_taylor_jeo} for our analysis. Choosing a very conservative limit, we assume that the FFT gives a correct answer for the power spectrum if $P_{\rm{FFT}}(k)/P_{\rm{FT}}(k)\lesssim0.5\%$: this holds for $k\approx 0.314 \,h\,\rm Mpc^{-1} = 67\%k_{\rm N}$, which is indicated by the vertical dotted line in Figure \ref{fig:testMAS}.

\bsp

\label{lastpage}

\end{document}